\documentclass[a4paper,UKenglish,cleveref, autoref, thm-restate]{lipics-v2021}
\hideLIPIcs

\usepackage{tikz}
\usetikzlibrary{positioning,arrows.meta,calc,shapes.symbols,shapes.arrows}

\title{From Formal Methods to Fuzzing and Back}

\author{Alastair F. Donaldson}{Department of Computing, Imperial College London, UK}{alastair.donaldson@imperial.ac.uk}{https://orcid.org/0000-0002-7448-7961}{}

\authorrunning{A.\,F. Donaldson}

\Copyright{Alastair F. Donaldson}



\category{Invited Talk}

\relatedversion{}

\nolinenumbers

\begin{document}

\maketitle

\begin{abstract}
  For the first twelve years of my career in computer science I mainly worked on \emph{formal methods}.
  I had been drawn to the field as an undergraduate by its promise of a way to build software with strong, mathematically based correctness guarantees.
  But after many years of struggling to make formal methods scale,
  I was seduced by randomised testing---also known as \emph{fuzzing}.
  Fuzzing is much less principled than formal methods,
  but it \emph{just works}:
  it can be readily applied to very large systems, and it is extremely effective at finding real bugs that developers care about and fix.
  In this personal account I trace my path from formal methods to fuzzing, by way of the GPUVerify project on verifying GPU kernels and the GraphicsFuzz project on randomised testing of GPU compilers.
  I argue that the \emph{oracle problem} provides an intellectual link between fuzzing and formal methods,
  describe my continuing, targeted uses of formal methods in the domain of GPU programming language design,
  and discuss work at the boundary between formal methods and fuzzing.
  Finally, I reflect on whether generative AI might be the missing ingredient that allows formal methods to scale,
  and set out my plans for large-scale formal methods via the Keystone project,
  through which my collaborators and I aim to find out how much of that promise can be realised in practice.
\end{abstract}

\section{Introduction}

I first encountered \emph{formal methods} in the penultimate year of my Bachelor's degree at the University of Glasgow,
through a course entitled ``Professional Software Development''.
Taught by the late Professor Ray Welland,
the course mainly focused on software engineering processes,
covering topics such as requirements elicitation,
project management,
legacy systems
and re-engineering.
While interesting and practical,
the material did not feel like core computer science:
little about it was mathematical or precise.
But at the very end of the course, Professor Welland gave two lectures about the Z notation~\cite{DBLP:conf/setss/Bowen14},
and I got a glimpse into something that seemed much more satisfying:
the idea of using a mathematical formalism to precisely specify properties of software.
Then, in my final year, I took
``Modelling Reactive Systems'',
a course on model checking
taught by Professor (and now Dame) Muffy Calder.
This was even better!
We learned how to describe a distributed protocol in a formalism very close to source code (Promela),
express desired properties of the protocol in logic (LTL),
feed both to the SPIN model checker~\cite{holzmann2004spin}
and---quick as a flash---get an execution trace demonstrating that the protocol did not conform to the specification
(indicating a bug in one or the other).
I was hooked. Upon graduating, I embarked on a PhD in formal methods.

I worked primarily on formal methods from 2003 to 2015 as a PhD student, postdoc and early career faculty member.
But although I learned a lot and had a lot of fun in the process,
there was always an elephant in the room:
the real-world impact of formal methods was very limited.
Despite islands of success (examples including Microsoft's Static Driver Verifier~\cite{DBLP:conf/ifm/BallCLR04},
the seL4 verified OS kernel~\cite{DBLP:conf/sosp/KleinEHACDEEKNSTW09} and the CompCert verified compiler~\cite{DBLP:journals/cacm/Leroy09}),
in general it turned out to be difficult to make formal methods \emph{scale}.
The lack of formal methods impact would occasionally be called out explicitly.
I recall a fellow PhD student at Glasgow
bluntly telling me: ``you know, nobody is \emph{ever} going to use this stuff''.
When I was a postdoc at Oxford, a colleague's advice was:
``if you want to have impact, change field'' (this colleague also worked on formal methods, and was at peace with the idea of not having impact).
As an early career faculty member at Imperial,
one of my PhD students was chided by another student (in jest, but almost accurately)
about there being nearly as many co-authors on one of our papers as there were lines of code in the program the paper sought to formally verify!

From around 2014 I started experimenting with randomised testing, also known as \emph{fuzzing}.
I was quickly seduced by the fact that fuzzing \emph{just works}.
I felt, first-hand, the thrill of being able to find real bugs in industry-scale software projects spanning millions of lines of code---bugs that
developers quickly confirmed and fixed.
This was the kind of impact I had been yearning for,
and over the last decade my work has mainly focused on software testing and fuzzing,
including founding a company, GraphicsFuzz Ltd., that was acquired by Google.

Still, despite the \emph{practical} satisfaction that fuzzing brings,
I find formal methods more \emph{fundamentally interesting}.
Fuzzing takes a ``suck it and see'' approach, where the worth of an idea is judged by its ability to find bugs.
But there is nothing fundamental about finding bugs: whether one finds bugs in a system under test (SUT) is as much a function of how reliable the SUT is (including the extent to which the SUT has already been hardened by previous fuzzing campaigns) as it is a function of how smart and rigorous the current fuzzing strategy is.
And ultimately, fuzzing provides no guarantees; Dijkstra famously stated the obvious fact:
``Program testing can be used to show the presence of bugs, but never to show their absence!''~\cite{dijkstra1970notes}.
So, I never fully left formal methods: I kept working on problems where scale is not a blocker, such as using formal methods to improve the rigour of GPU programming language semantics, as well as working at the intersection of the two fields, using formal methods as a basis for randomised testing, and bringing fuzzing techniques to bear on formal methods tools.

But now, in 2026, it seems that formal methods might finally have their day, thanks to the artificial intelligence revolution.
Large language models (LLMs) are shockingly good at producing the mathematical artifacts---such as invariants and proofs---that formal methods demand and that humans struggle to grapple with.
Furthermore, they can produce these directly in formal notation,
e.g.\ writing proofs for proof assistants such as Lean and Rocq.
And, better still, the outputs of LLMs do not need to be trusted when used in many formal methods contexts:
they can be independently and rigorously checked by (non-AI-based) proof checkers---a marriage made in heaven!
While LLM capabilities have ballooned month-on-month over the last four years,
my experience is that the frontier models of late 2025 (for me, Claude Opus 4.5) marked a step change in capability with respect to formal proof;
enough of a step change that formal methods projects that would traditionally have required heroic efforts that few would dare undertake might soon become mainstream.
My first direct taste of this came via a project where I used Lean (with which I had no prior experience) to prove a result connecting randomised generation of structured data with the languages accepted by Turing machines~\cite{DBLP:journals/corr/abs-2607-11811}.
Despite my lack of prior experience, AI enabled me to make very rapid progress on what turned out to be a sizeable proof~\cite{yu2026randomisedformalisation}.
Inspired by this, I have recently embarked on \emph{Keystone}~\cite{aria2026keystone}, a collaboration between Imperial College London and EPFL, funded by the UK's Advanced Research and Invention Agency (ARIA), on using AI and proof assistants to construct formally verified software, with a particular focus on creating formally verified infrastructure for serving LLMs.

This article is a personal account and reflection on my experience working in and moving between the formal methods and software testing fields.
The examples I give, and the formal methods and fuzzing techniques I discuss,
are thus heavily influenced by my own experience.
Throughout, I consider formal methods applied to \emph{software}, though some of the points made apply equally well to hardware verification, and some of the remarks about scaling proofs may be relevant to the use of proof assistants in mathematics.

In the rest of this article I elaborate on:

\begin{itemize}

\item the scalability challenges that have hindered the practical impact of formal methods (\Cref{sec:scale});

\item my experience wrestling with these challenges in the GPUVerify project on formal verification for GPU kernels (\Cref{sec:gpuverify});

\item my move from formal methods to fuzzing, which led to the GraphicsFuzz project, with some thoughts on why the \emph{oracle problem} provides an intellectual link between the two fields (\Cref{sec:graphicsfuzz});

\item my continued work on ``small scale'' formal methods, in the context of GPU programming language semantics (\Cref{sec:smallscale});

\item my work at the intersection of formal methods and fuzzing (\Cref{sec:fmfuzzing});

\item the opportunities that the AI revolution brings for formal methods, and the Keystone project through which my collaborators and I plan to explore them (\Cref{sec:ai}).

\end{itemize}

If your background brings to mind different examples or experiences that either support or run counter to the points I make in this article, I would love to hear from you.

\section{Challenges of Scale}\label{sec:scale}

Why have formal methods struggled to have practical impact?
In my experience the answer almost always comes down to \emph{scale},
but ``scale'' covers several distinct problems.
Broadly, I identify four main challenges of scale:
the scale of proofs,
the scale at which automated verification algorithms can operate,
the scale of the tooling needed to handle real-world software,
and the scale of teams that work on formal methods projects.

I describe these problems of scale as they stood before the recent advances in AI.
Some of them look rather different now, and I return to them in \Cref{sec:ai}.

\paragraph*{Proof scale}
Interactive theorem provers such as Isabelle~\cite{DBLP:books/sp/NipkowPW02}, Lean~\cite{DBLP:conf/cade/Moura021} and Rocq~\cite{Rocq} (formerly Coq) sit at one end of the formal methods spectrum---the ``highly manual'' end.
Their big advantage is that,
if you are willing to put the manual work in,
you can express the argument needed to prove a program correct in as intricate a fashion as is required.
The only limits are (a) how elaborate an argument must be communicated to the system for a proof to go through,
and (b) how much time the user is able to invest in the process of discovering and communicating a detailed argument.

The problem is that the argument that must be communicated to the system can be \emph{very}
elaborate.
For example, the authors of the seL4 formally verified kernel reported that while the kernel comprises 8,700 lines of C,
its refinement proofs comprise around 165,000 lines of Isabelle,
and the complete development, including frameworks, libraries and generated proofs, comes to some 200,000 lines~\cite{DBLP:conf/sosp/KleinEHACDEEKNSTW09}.
CompCert tells a similar story:
of the development's 42,000 lines of Coq,
around 6,000 are the compiler itself and around 4,000 are the semantics of the source, intermediate and target languages,
while the remaining 32,000 are the correctness proof,
so that there are more than five lines of proof for every line of compiler code~\cite{DBLP:journals/cacm/Leroy09}.
And an empirical study of the seL4 proofs, together with the largest software verification developments in the Archive of Formal Proofs, suggests that the size of a proof grows quadratically with the size of the statement being proved~\cite{DBLP:conf/icse/MatichukMAJKS15}.

Another problem is that when working with a proof assistant one never knows how close one is to the end,
and there are always twists and turns.
For example, a leaf lemma may turn out not to be provable without some extra piece of machinery that
then needs to be threaded through every theorem that depends (directly or indirectly) on the lemma,
or a subtle flaw in a definition or theorem statement can turn out to make a result untrue as stated,
requiring major proof surgery.
A related problem is that the effort required to update a proof when changes are made to the associated source code
can be substantial:
the seL4 team reported that adding \emph{reply capabilities} to the kernel required substantial updates to existing proofs,
with re-verification taking around a person-year~\cite{DBLP:conf/sosp/KleinEHACDEEKNSTW09}.
DeMillo, Lipton and Perlis anticipated this kind of problem decades earlier in their 1979 critique of formal verification~\cite[p.~278]{DBLP:journals/cacm/DeMilloLP79}:

\begin{quote}
  ``The discontinuous nature of programming sounds the death knell for verification. A sufficiently fanatical researcher might be willing to devote two or three years to verifying a significant piece of software if he could be assured that the software would remain stable. But real-life programs need to be maintained and modified. There is no reason to believe that verifying a modified program is any easier than verifying the original the first time around.''
\end{quote}

\paragraph*{Algorithmic scale}
At a different point on the formal methods spectrum are verification techniques that aim to be fully or highly automatic,
such as model checking.
Here the limits of scale are algorithmic.
For software with unbounded state (integers, dynamically allocated memory, recursion), most interesting verification problems are undecidable, by reduction from the halting problem.
Even when restricted to finite-state systems, where the problem is decidable, the state space of a system typically grows exponentially in the number of its components,
so that exhaustive exploration quickly becomes infeasible:
this is the infamous \emph{state explosion problem}~\cite{DBLP:conf/laser/ClarkeKNZ11}.

Key categories of techniques that respond to the problem of algorithmic scale include \emph{abstraction}~\cite{DBLP:journals/toplas/ClarkeGL94,DBLP:conf/pldi/BallMMR01,DBLP:conf/popl/HenzingerJMS02},
which shrinks the state space of a system by discarding detail,
\emph{pruning techniques} (e.g.\ partial order reduction~\cite{DBLP:books/sp/Godefroid96} and symmetry reduction~\cite{DBLP:journals/fmsd/ClarkeJEF96,DBLP:journals/fmsd/EmersonS96}), which reduce the algorithmic work required by avoiding the need to explore
parts of the state space that are provably equivalent to others,
and \emph{symbolic methods}, which represent sets of states implicitly (typically using BDDs~\cite{DBLP:journals/iandc/BurchCMDH92}, SAT formulas~\cite{DBLP:conf/tacas/BiereCCZ99} or SMT formulas~\cite{DBLP:journals/sttt/ArmandoMP09,DBLP:journals/tse/CordeiroFM12}).

While these techniques have dramatically improved the scalability of algorithmic verification techniques, they all come with limitations and trade-offs.
This is unavoidable since the underlying problems are intractable or undecidable in general.
Abstraction leads to over-approximate verification techniques that are sound but may report false alarms.
Abstraction refinement, which uses spurious counterexamples to recover just enough detail to rule them out~\cite{DBLP:conf/cav/ClarkeGJLV00},
is great when it works.
However, the refinement loop can diverge,
generating ever more refinements without converging on a proof~\cite{DBLP:conf/tacas/JhalaM06}.
State-space pruning techniques can achieve dramatic reductions on systems with the right structure,
but they do not fundamentally solve the problem of scale.
While symmetry reduction can give a factorial saving when applied to a distributed protocol over a fully connected topology,
the saving shrinks as the degree of symmetry falls.
A ring topology, for instance, yields only a linear reduction,
and for systems with no symmetry at all,
symmetry reduction offers no benefit.
Partial-order reduction collapses commuting interleavings in a concurrent system, but the reduced space is still typically exponential in the number of concurrent components.
And while symbolic methods allow verification to scale by orders of magnitude compared with explicit-state representations,
the underlying techniques can be brittle (for instance, the size of a BDD can vary dramatically depending on how variables are ordered~\cite{DBLP:journals/tc/Bryant86}), making them highly sensitive to how things are modelled.

The brittle nature of symbolic techniques is a particular problem in so-called \emph{auto-active} verification~\cite{leino2010autoactive},
which sits between fully automatic model checking and interactive theorem proving.
In an auto-active verifier such as Dafny~\cite{DBLP:conf/lpar/Leino10}, Why3~\cite{DBLP:conf/esop/FilliatreP13}, Viper~\cite{DBLP:conf/vmcai/0001SS16} or F$^\star$~\cite{DBLP:conf/popl/SwamyHKRDFBFSKZ16},
the user annotates the program with specifications (pre- and post-conditions, loop invariants, assertions and lemmas) and the tool generates verification conditions from the annotated program and discharges them automatically by calling an SMT solver.
While leveraging SMT solvers can improve automation (compared with interactive theorem proving),
the brittle nature of solvers leads to fragility.
Verification conditions often involve quantifiers,
and SMT solvers handle quantified formulas by heuristic instantiation
(e.g.\ via \emph{E-matching} against syntactic trigger patterns~\cite{DBLP:conf/cade/MouraB07}),
which is incomplete and the behaviour of which depends on the precise shape of the query.
Because the user has only indirect control over that shape, through annotations whose effect on the solver is opaque,
a failed proof attempt offers little guidance as to what to change.
As a result, SMT-based program verifiers suffer from \emph{instability},
whereby a semantically irrelevant change to a query, such as renaming identifiers or reordering assertions,
can cause a previously successful proof to fail or time out~\cite{DBLP:conf/fmcad/ZhouBTLHP23}.
Amazon Web Services, which has used Dafny heavily and contributed to its development, has identified verification instability as a factor that hindered the adoption of Dafny,
reporting that production projects at AWS see previously-verified code stop verifying after small, seemingly unrelated edits~\cite{salkeld2022cost},
and has published guidance on structuring Dafny code to reduce the risk~\cite{tomb2023brittleness};
engineers on
AWS's formally verified authorization engine report that brittleness grew with the codebase to the point of delaying deployments~\cite{DBLP:conf/icse/ChakarovGHHHJJL25}.

\paragraph*{Tool scale}
Getting practitioners interested in a technique requires giving them something they can try out on their own code, in their own domain.
Yet formal methods are often defined and evaluated in a simplified setting such as a core calculus or toy language.
Making them work for real-world languages---with gnarly features such as preprocessor macros, templates, foreign-function interfaces, inline assembly and a rich infrastructure of libraries managed by build systems---is very difficult from a practical point of view.
DeMillo, Lipton and Perlis make this point vividly,
characterising real production systems through the programmer's jargon of
``[p]atches, ad hoc constructions, bandaids and tourniquets, bells and whistles, glue, spit and polish''~\cite[p.~277]{DBLP:journals/cacm/DeMilloLP79}.
Scaling formal methods to real languages requires substantial tooling work that formal methods researchers may not want to do or know how to do.
Furthermore, undertaking such tooling work may be high risk from a career advancement point of view given that efforts towards building tools are not valued as highly in academic computer science as I believe they should be.

Various industry reports on applying static analysis techniques to large codebases
(varying in the extent to which the associated methods would be deemed ``formal'')
bear out the need for attention to tooling issues. The team behind the Coverity static analyser found that parsing the many dialects of C accepted by their customers' compilers,
and fitting into customers' build processes,
were major obstacles that had to be overcome before any analysis could take place~\cite[pp.~69--70]{DBLP:journals/cacm/BesseyBCCFHHKME10};
reports from Google and Facebook similarly emphasise the importance of integration into developer workflows~\cite{DBLP:journals/cacm/SadowskiAEMJ18,DBLP:journals/cacm/DistefanoFLO19}.

The availability of open-source program analysis frameworks has helped a great deal in this regard,
e.g.\ the Clang/LLVM project~\cite{DBLP:conf/cgo/LattnerA04} provides industrial-strength front ends for C-family languages;
the CProver framework that underpins CBMC~\cite{DBLP:conf/tacas/ClarkeKL04},
and configurable program analysis~\cite{DBLP:conf/cav/BeyerHT07} as realised in the CPAchecker framework~\cite{DBLP:conf/cav/BeyerK11},
give verification researchers shared infrastructure to build on;
and intermediate verification languages such as Boogie~\cite{DBLP:conf/fmco/BarnettCDJL05}
play an analogous role in the development of verification tools as compiler intermediate representations play in the development of compilers, avoiding the need to build a verification tool entirely from scratch.
GPUVerify, which I discuss in \Cref{sec:gpuverify}, relied on both Clang/LLVM and Boogie.
But while such frameworks lower the barrier to building formal methods tools,
their main benefit is to take away the pain of parsing and type-checking a real language.
Developers of a tool built on such a framework still need to decide how to handle the weird and wonderful features that the front end faithfully passes on.

\paragraph*{Team scale}
To bring a formal method to bear on a project, one first needs to build a team whose members have mastery of the formal method,
and gaining mastery of a formal method can be demanding.
One of Hall's ``Seven Myths of Formal Methods'' is ``Formal methods require highly trained mathematicians''~\cite[Myth 4]{DBLP:journals/software/Hall90}.
However, Hall's argument that this is a myth mainly relates to writing specifications: ``The fact is that the mathematics for specification is easy. Once it is recognized that the practice of formal methods is most concerned with writing specifications, the mathematical difficulties become much less significant. You can develop specifications themselves with very straightforward mathematics that any practicing engineer should know''~\cite[p.~16]{DBLP:journals/software/Hall90}.
My experience is both that writing specifications is more challenging than Hall states,
and that using formal methods tools tends to involve a lot more than specification writing:
in particular, many formal methods involve writing proofs, which can be a major challenge.
The steep learning curve associated with interactive theorem provers is well known~\cite{DBLP:conf/iwpc/JuhosovaZC25,DBLP:journals/ftpl/RingerPSGT19}.

Beyond the challenge of building a team, there is the problem of
scaling formal methods work \emph{across} a team.
Agile software development revolves around small, independent increments that can be delivered quickly.
Agile \emph{formal methods} work is harder to achieve,
because a small change to a codebase may necessitate a substantial change to a proof,
or may derail an automated verification technique by upsetting the conditions for which its heuristics have been carefully tuned.
An experience report by members of the seL4 project acknowledges this challenge:
``Ideally [distribution of proof sub-tasks to team members] is achieved by a compositional
calculus that allows clean, small interfaces between sub-tasks, which are defined
once and remain stable. In reality, this is rarely the case. Full compositionality
often comes with a high price in other aspects of complexity, while interfaces are
rarely small or stable.''~\cite[p.~40]{DBLP:conf/aisc/BourkeDKK12}
Nevertheless, with a suitably designed proof structure, the seL4 team were able to divide proof work among team members:
``the FM [formal methods] team managed to design refinement calculi for each of the major proof steps that allowed the work to be divided up among members of the team''~\cite[p.~1008]{DBLP:conf/icse/AndronickJKKSZZ12}.

\section{GPUVerify: My Formal Methods Push}\label{sec:gpuverify}

My most sustained attempt to make formal verification work in practice was GPUVerify,
a static verification tool for GPU kernels written in OpenCL and CUDA,
developed with a team of PhD students, postdocs and collaborators at Imperial College London and Microsoft Research~\cite{DBLP:journals/toplas/BettsCDKQTW15,DBLP:conf/cav/BardsleyBCCDDKLQ14}.
A GPU kernel is a function executed simultaneously by thousands of threads,
and a recurring source of bugs in such kernels is the \emph{data race},
where two threads access a common memory location, at least one of the accesses is a write,
and the accesses are not ordered by inter-thread synchronisation.
Our hope with GPUVerify was that
by targeting a niche-but-important problem---proving race-freedom for GPU kernels, rather than proving arbitrary functional properties for arbitrary software---we
could build a highly automatic formal methods tool~\cite{DBLP:conf/cav/BardsleyBCCDDKLQ14}.
In terms of the challenges of \Cref{sec:scale},
we addressed tool scale by building on existing infrastructure, as I describe below;
the tightly scoped problem meant that a relatively small team sufficed;
and our focus on automation was a deliberate attempt to sidestep proof scale.
That left algorithmic scale, and in particular the fragility of SMT-based verification, as the challenge we could not avoid.

\begin{figure}
\resizebox{\textwidth}{!}{
  \begin{tikzpicture}[
  every node/.style={align=center, minimum height=9mm, inner sep=3pt},
  data/.style={draw, fill=black!15},
  tool/.style={draw, rounded corners=2pt, fill=white},
  arr/.style={-{Latex[length=2.2mm]}, thick},
]
  \node[data, minimum width=17mm] (opencl) at (0,0.65) {OpenCL\\kernel};
  \node[data, minimum width=17mm] (cuda)   at (0,-0.65) {CUDA\\kernel};

  \node[tool, right=6mm of $(opencl.east)!0.5!(cuda.east)$] (frontend) {Front-end: built\\on Clang/LLVM};
  \node[tool, right=6mm of frontend] (kte) {Kernel\\transformation\\engine};

  \node[data, anchor=west, minimum width=31mm] (boogieprog) at ($(kte.east)+(8mm,7mm)$) {Sequential Boogie\\program};
  \node[data, anchor=west, minimum width=31mm] (invariants) at ($(kte.east)+(8mm,-7mm)$) {Candidate loop\\invariants};

  \node[tool, right=8mm of $(boogieprog.east)!0.5!(invariants.east)$] (engine) {Boogie\\verification\\engine};
  \node[tool, right=6mm of engine] (smt) {SMT\\solver};

  \draw[arr] (opencl.east) -- ($(frontend.west)+(0,2mm)$);
  \draw[arr] (cuda.east)   -- ($(frontend.west)+(0,-2mm)$);
  \draw[arr] (frontend) -- (kte);
  \draw[arr] (kte.east) -- (boogieprog.west);
  \draw[arr] (kte.east) -- (invariants.west);
  \draw[arr] (boogieprog.east) -- ($(engine.west)+(0,2mm)$);
  \draw[arr] (invariants.east) -- ($(engine.west)+(0,-2mm)$);
  \draw[{Latex[length=2.2mm]}-{Latex[length=2.2mm]}, thick] (engine) -- (smt);
  \end{tikzpicture}
  }
\caption{The components comprising the GPUVerify tool for verifying data race-freedom of GPU kernels, adapted from~\cite{DBLP:journals/eceasst/Donaldson14}}\label{fig:gpuverifyoverview}
\end{figure}
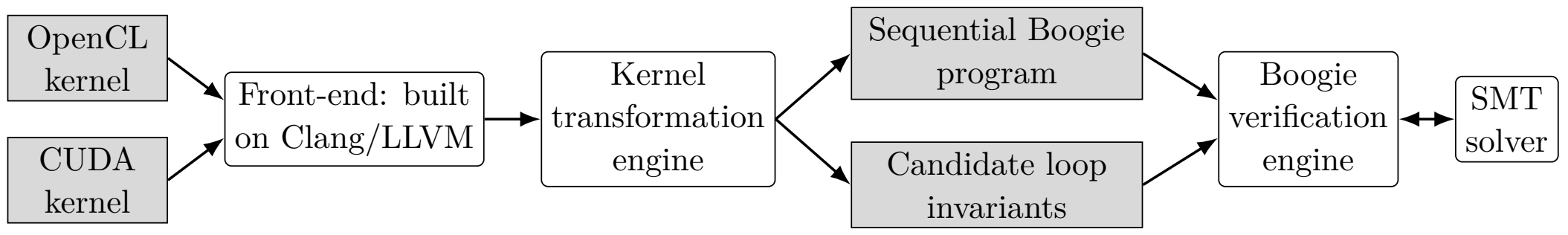

\paragraph*{GPUVerify: a brief overview}
The main insight behind GPUVerify is that if
execution barriers are the only means of synchronisation (a property of many GPU kernels, and one that was almost ubiquitous when we started the project in 2011),
a race-free kernel behaves deterministically.
Thus, as long as races are detected, it suffices to analyse a single, canonical thread schedule.
This allows a massively parallel kernel to be rewritten as a sequential program.
On its own this does not scale, because the sequential program would encode the work of thousands of threads and would thus be infeasibly large.
GPUVerify therefore applies a \emph{two-thread abstraction}:
since a data race is a conflict between exactly two threads,
it suffices to check an arbitrary pair of distinct threads,
with the effects of all other threads over-approximated by treating shared memory abstractly.
Combined with predicated execution to handle thread-dependent control flow,
and instrumentation to log and check memory accesses,
GPUVerify turns a kernel into a sequential program,
expressed in the Boogie intermediate verification language~\cite{DBLP:conf/fmco/BarnettCDJL05}.
Correctness of this sequential program implies that the original kernel is free from data races (and from barrier divergence, a related defect).
Verifying the sequential program requires loop invariants,
which GPUVerify infers using the Houdini algorithm~\cite{DBLP:conf/fm/FlanaganL01}:
a large set of \emph{candidate} invariants is guessed,
using domain-specific rules that capture memory access idioms we observed in real kernels,
and Houdini repeatedly discards candidates that cannot be proven until the survivors form an inductive invariant.
The components that comprise GPUVerify are illustrated by \Cref{fig:gpuverifyoverview}.

\paragraph*{Standing on the shoulders of giants}
GPUVerify would not have been possible without heavy re-use of existing infrastructure.
We benefitted from OpenCL and CUDA support in the Clang/LLVM framework, which handled the parsing and lowering of kernels to LLVM IR,
sparing us from dealing with the full syntactic complexity of the source languages
(C++ templates in CUDA code being a case in point).
Our own translator then mapped LLVM IR into Boogie,
and Boogie, in turn, handled verification condition generation and invariant inference,
delegating to the Z3~\cite{DBLP:conf/tacas/MouraB08} and CVC4~\cite{DBLP:conf/cav/BarrettCDHJKRT11} SMT solvers.
This allowed us to apply GPUVerify directly to developer-written kernels, and we evaluated the tool on
all the publicly available GPU benchmark suites we could get our hands on at the time,
discovering several real data races along the way.
An industrial collaborator, Rightware (a Finnish company specialising in benchmarking), used GPUVerify to verify race-freedom across its Basemark CL benchmark suite, finding a defect in the process~\cite{DBLP:conf/cav/BardsleyBCCDDKLQ14}.
 
\paragraph*{GPUVerify: limits of scale}
Despite its relative success on a number of benchmark suites, GPUVerify also ran into some of the problems of scale described in \Cref{sec:scale},
principally the problems associated with auto-active verifiers discussed at the end of ``Algorithmic scale''.

First, invariant inference was unpredictable.
Verifying kernels with loops typically required intricate invariants
relating the memory locations that a thread accesses to its thread and block ids:
invariants we could not reasonably expect GPU programmers to write.
Our domain-specific candidate-generation rules let Houdini find such invariants for many kernels~\cite{DBLP:journals/tse/BettsCDDK18}.
This often worked out of the box, but when it did not the user was left high and dry,
facing a possible data race report that was indistinguishable at first glance from a real bug.
Speculating more candidates also made verification slower and its running time harder to predict.
Second, we were at the mercy of the quirks of SMT solvers.
Much of our engineering effort went into solver-friendly encodings,
such as quantifier-free representations of the sets of memory accesses used for race checking.
We found ourselves talking to solver developers to find just the right solver options to use,
and there was no guarantee that our encodings would remain efficient as the solvers evolved.
Performance was excellent for some kernels but miserable for others,
switching between Z3 and CVC4 could change the picture dramatically,
and even an optimisation that simplified the program to be verified could make some kernels much slower to verify,
often for no apparent reason~\cite{DBLP:conf/cav/BardsleyBCCDDKLQ14}.
Third, the two-thread abstraction that made GPUVerify scale also imposed a hard limit on what it could prove.
Because the effects of all other threads on shared memory are over-approximated,
GPUVerify reports false alarms for \emph{data-dependent} kernels,
where race-freedom depends on values that other threads have written to shared memory.
We introduced barrier invariants, a richer shared state abstraction, to handle some of these cases~\cite{DBLP:conf/oopsla/ChongDKKQ13},
but they needed to be written by hand.

In practice, data-dependent kernels were a minority,
so the two-thread abstraction was a good trade-off due to its scalability benefits.
My overall feeling about GPUVerify is that it scaled well enough, in the sense of handling the majority of kernels from real benchmark suites,
but that it owed this to a domain-specific abstraction rather than to a more general advance in verification technology.
What ultimately limited the tool was not the size of the kernels but the unpredictability of the automated reasoning on which it relied, for invariant generation and SMT solving alike.

\section{GraphicsFuzz: From Formal Methods to Fuzzing}\label{sec:graphicsfuzz}

When I gave talks about GPUVerify,
a question I was often asked was:
``but can you trust the GPU compiler?''
Proving that a kernel is free from data races is of limited comfort
if the compiler that translates it into machine code introduces bugs of its own.
Because GPU kernels are compiled at runtime by the driver of whatever GPU happens to be present,
an application developer cannot easily anticipate or work around such bugs.
My friend Anton Lokhmotov,
then at ARM and a collaborator on the CARP EU project,
pointed me to Csmith~\cite{DBLP:conf/pldi/YangCER11},
a tool that generates random C programs and cross-checks C compilers against one another,
and which had found hundreds of bugs in GCC and LLVM.
``Maybe you should look at doing something similar for OpenCL,'' he said.
I recruited a talented undergraduate, Christopher Lidbury,
who built CLsmith, an OpenCL adaptation of Csmith~\cite{DBLP:conf/pldi/LidburyLCD15}.
It proved tremendously good at finding bugs:
more than 50 in OpenCL compilers, most of them in commercial implementations.
Coming from several years working on GPUVerify, this was a revelation:
there were no false alarms to explain away
and no scalability problems to wrestle with;
the tool readily found real bugs that developers cared about and fixed.

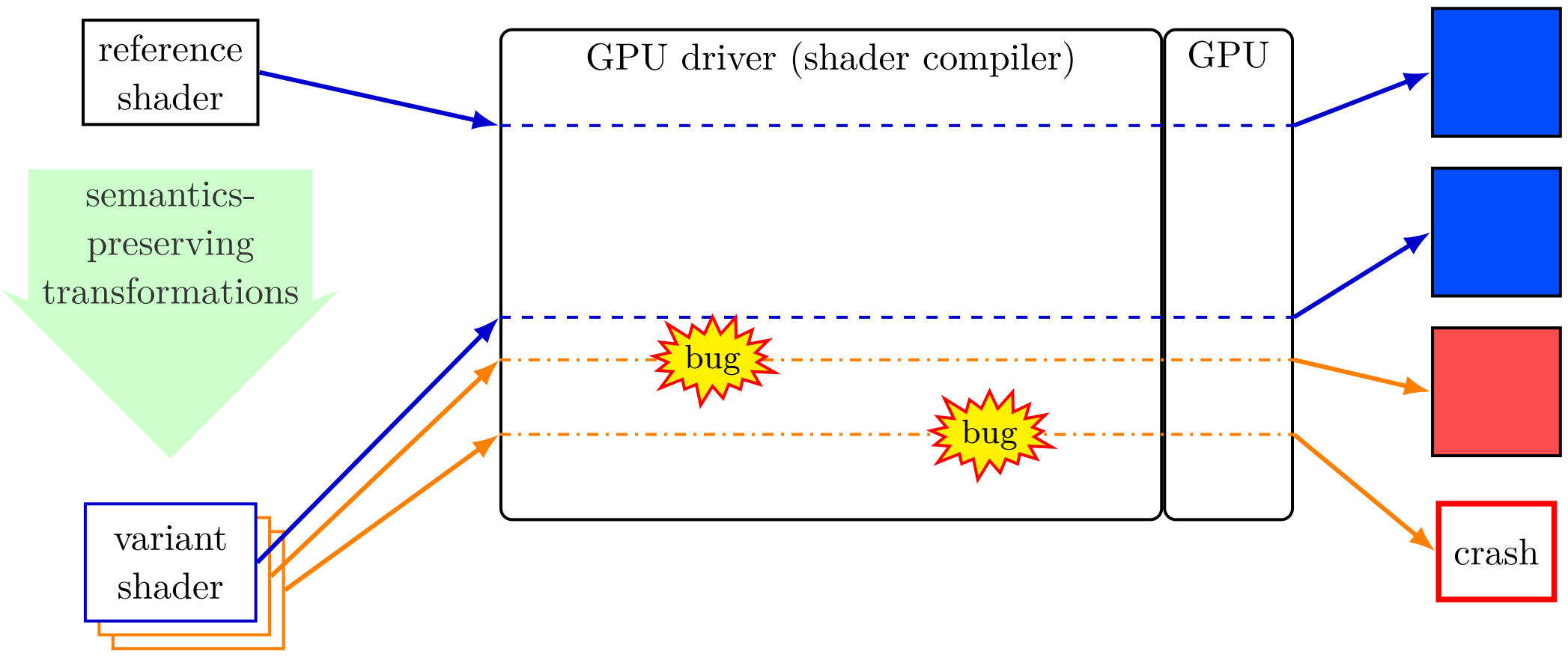
\begin{figure}
\resizebox{\textwidth}{!}{
\begin{tikzpicture}[
  box/.style={draw, thick, fill=white, align=center, inner sep=4pt, minimum height=9mm},
  bug/.style={starburst, draw=red, thick, fill=yellow, inner sep=1pt, font=\small, starburst point height=2.2mm},
  img/.style={minimum width=12mm, minimum height=12mm, inner sep=0pt, draw, thick},
  flow/.style={thick},
  arr/.style={-{Latex[length=2.5mm]}, line width=1.2pt},
  refcol/.style={color=blue!80!black},
  badcol/.style={color=orange},
]
  \node[box] (ref) at (0,0) {reference\\shader};
  \node[single arrow, shape border rotate=270, draw=none, fill=green!20,
        minimum height=20mm, minimum width=22mm, single arrow head extend=2.5mm, single arrow head indent=1mm,
        below=4mm of ref, align=center, text=black!80]
        (xform) {semantics-\\preserving\\transformations};
  \node[box, draw=orange, minimum width=1.6cm, minimum height=1.1cm,
        below=4mm of xform, xshift=2.6mm, yshift=-2.6mm] (var3) {};
  \node[box, draw=orange, minimum width=1.6cm, minimum height=1.1cm,
        below=4mm of xform, xshift=1.3mm, yshift=-1.3mm] (var2) {};
  \node[box, draw=blue!80!black, minimum width=1.6cm, minimum height=1.1cm,
        below=4mm of xform] (var) {variant\\shader};

  \node[draw, thick, rounded corners=3pt, minimum width=62mm, minimum height=46mm]
        (driver) at (6.2,-1.9) {};
  \node[anchor=north] at (driver.north) {GPU driver (shader compiler)};
  \node[draw, thick, rounded corners=3pt, minimum width=12mm, minimum height=46mm, right=0mm of driver] (gpu) {};
  \node[anchor=north] at (gpu.north) {GPU};

  \coordinate (L) at (driver.west);
  \coordinate (R) at (gpu.east);
  \draw[flow, refcol, dashed]   ($(L)+(0,1.4)$)  -- ($(R)+(0,1.4)$);
  \draw[flow, refcol, dashed]   ($(L)+(0,-0.4)$) -- ($(R)+(0,-0.4)$);
  \draw[flow, badcol, dash dot] ($(L)+(0,-0.8)$) -- ($(R)+(0,-0.8)$);
  \draw[flow, badcol, dash dot] ($(L)+(0,-1.5)$) -- ($(R)+(0,-1.5)$);

  \node[bug] at ($(L)+(2.0,-0.8)$) {bug};
  \node[bug] at ($(L)+(4.6,-1.5)$) {bug};

  \node[img, fill=blue!70!cyan, draw=black] (o1) at ($(R)+(1.9,1.9)$) {};
  \node[img, fill=blue!70!cyan, draw=black] (o2) at ($(R)+(1.9,0.4)$) {};
  \node[img, fill=red!70,       draw=black] (o3) at ($(R)+(1.9,-1.1)$) {};
  \node[box, draw=red, line width=1.5pt] (o4) at ($(R)+(1.9,-2.6)$) {crash};

  \draw[arr, refcol] (ref.east)  -- ($(L)+(0,1.4)$);
  \draw[arr, refcol] (var.east)  -- ($(L)+(0,-0.4)$);
  \draw[arr, badcol] (var2.east) -- ($(L)+(0,-0.8)$);
  \draw[arr, badcol] (var3.east) -- ($(L)+(0,-1.5)$);

  \draw[arr, refcol] ($(R)+(0,1.4)$)  -- (o1.west);
  \draw[arr, refcol] ($(R)+(0,-0.4)$) -- (o2.west);
  \draw[arr, badcol] ($(R)+(0,-0.8)$) -- (o3.west);
  \draw[arr, badcol] ($(R)+(0,-1.5)$) -- (o4.west);
\end{tikzpicture}
}
\caption{An overview of the GraphicsFuzz approach to testing GPU shader compilers, adapted from~\cite{DBLP:conf/ecoop/DonaldsonET19}}\label{fig:graphicsfuzz}
\end{figure}

\paragraph*{The oracle problem for graphics compilers}
I next turned to compilers for graphics shading languages such as GLSL, the OpenGL shading language,
initially because they posed an intriguing puzzle.
Shading languages are deliberately underspecified with respect to floating-point precision,
so as to cater for a diverse range of current and future GPUs.
As a result there is no single correct image that a shader should render~\cite{DBLP:journals/pacmpl/DonaldsonELT17}.
As someone from a formal methods background I found this fascinating:
you clearly cannot formally verify a compiler against such a loose specification,
but can you reliably test it?
This led me to Weyuker's classic work on testing ``non-testable'' programs,
whose correct output is not known in advance~\cite{DBLP:journals/cj/Weyuker82},
and to \emph{metamorphic testing}~\cite{MetamorphicTesting},
where instead of checking individual outputs,
one checks that related inputs yield appropriately related outputs.
I was particularly inspired by work on testing compilers via \emph{equivalence modulo inputs} (EMI)~\cite{DBLP:conf/pldi/LeAS14},
which mutates the parts of a program that are not executed on a given input,
yielding variants that should behave identically on that input.
We had already adapted the EMI idea to OpenCL by injecting code that is dead by construction~\cite{DBLP:conf/pldi/LidburyLCD15},
and for GLSL we generalised it to consider a range of semantics-preserving transformations~\cite{DBLP:journals/pacmpl/DonaldsonELT17}.
Our approach, implemented in a tool originally called GLFuzz and later renamed GraphicsFuzz, would create a set of \emph{variant} shaders from an original \emph{reference} shader. Each variant was produced through the application of a random sequence of semantics-preserving transformations.
By construction, a shader and its transformed variants should render visually identical images,
with a significantly different image being indicative of a compiler bug.
Crucially, because the reference and all variants would be executed on the same device, the compiler under test was only ever cross-checked against itself,
so that the problem of differing floating-point behaviour across GPUs became irrelevant.
We found that a straightforward image comparison metric was sufficient to tolerate small rounding differences arising due to optimisations, while still flagging up major differences arising from miscompilations.
\Cref{fig:graphicsfuzz} illustrates the GraphicsFuzz approach.

\paragraph*{From research to production}
In practice,
GraphicsFuzz
found bugs affecting every GPU and driver configuration we tested,
spanning all the major GPU designers~\cite{DBLP:journals/pacmpl/DonaldsonELT17}.
I had never had so much fun finding bugs.
Some of them were serious security vulnerabilities:
a malicious WebGL page viewed in Chrome on a Samsung Galaxy S6 could capture the contents of other browser tabs;
another could trigger a ``blue screen of death'' under Windows 10,
simply by being visited;
and further issues led to Apple fixing a memory corruption bug in iOS (CVE-2017-2424)
and NVIDIA fixing a machine freeze in its Linux drivers (CVE-2017-6259)~\cite{DBLP:journals/pacmpl/DonaldsonELT17}.

GraphicsFuzz formed the basis of a spin-out company, GraphicsFuzz Ltd.,
which Google acquired in 2018.
At Google, we put the technique into production to improve the Khronos Vulkan Conformance Test Suite (CTS),
which Vulkan drivers must pass in order to be deemed conformant,
and which is incorporated into the Android Compatibility Test Suite~\cite{DBLP:conf/ecoop/DonaldsonET19}.
The idea was simple:
whenever GraphicsFuzz finds a bug in a conformant driver,
this reveals a gap in the CTS,
so we built tooling to automatically find such bugs,
reduce the bug-inducing shaders to small, readable examples,
and package them as CTS tests,
ensuring that future conformant drivers cannot exhibit those bugs.
We also used differential coverage analysis to synthesise tests covering driver code that the CTS did not reach,
leading to hundreds of new CTS tests.

\paragraph*{The appeal of the oracle problem to a formal methods mindset}
Although fuzzing is where my work has had the most practical impact,
my interest in it is, at heart, that of a formal methods researcher.
What makes compiler fuzzing intellectually satisfying to me is the \emph{oracle problem}:
it is trivial to generate test inputs at random in a naive fashion,
but generating well-formed test programs for which a compiler should produce meaningful results is challenging,
and then knowing whether the output the compiler has generated is correct is hard.
Attacking the oracle problem calls for the kind of semantic reasoning that formal methods are built on:
metamorphic relations are, in effect, partial specifications,
the argument that a particular class of program transformations is semantics-preserving can be seen as a small theorem about the programming language of interest,
and creating a generator of programs that are free from undefined behaviour requires attention to the details of programming language semantics.

I find the link between formal methods and fuzzing provided by the oracle problem deeply appealing,
and some of my favourite randomised testing papers of recent years feature clever oracles.
For example, \emph{ternary logic partitioning}~\cite{DBLP:journals/pacmpl/RiggerS20} introduces an oracle for testing database engines
by exploiting the fact that, in SQL's three-valued logic,
every row satisfies exactly one of a predicate $p$, its negation $\neg p$, or $p~\texttt{IS NULL}$.
Thus the rows returned by a query such as \texttt{SELECT * FROM t}
should be \emph{exactly} the union of those returned by the three partitioned queries that add constraints such as
\texttt{WHERE c > 5}, \texttt{WHERE NOT (c > 5)} and \texttt{WHERE (c > 5) IS NULL}, respectively.
Any discrepancy between the union of the returned sets and the result returned by the unconstrained query reveals a bug.
Importantly, one does not need to know what the correct set of returned results is to know that there is a bug---this is the beauty of metamorphic testing.

Another example of a metamorphic oracle is \emph{semantic
fusion}~\cite{DBLP:conf/pldi/WintererZS20}, which tests SMT solvers
by fusing two formulas whose satisfiability is already known and
which are known to agree (both satisfiable or both unsatisfiable).
Using an illustrative example from the semantic fusion paper,
consider the satisfiable formulas
$\varphi_1 = x > 0 \wedge x > 1$ and $\varphi_2 = y < 0 \wedge y < 1$,
over disjoint variables.
Their conjunction is obviously satisfiable. To fuse them, one introduces a
fresh variable $z$ and a fusion constraint $z = x + y$ and replaces
some occurrences of $x$ by $z - y$ and of $y$ by $z - x$, yielding e.g.:
\[
  \varphi_{\mathrm{fused}} =
  (x > 0 \wedge z - y > 1) \wedge (z - x < 0 \wedge y < 1).
\]
This formula differs significantly from both seed formulas,
and it is satisfiable
by construction: any pair of models of $\varphi_1$ and $\varphi_2$
extends to a model of $\varphi_{\mathrm{fused}}$ by setting
$z = x + y$.
Thus if a solver reports $\varphi_{\mathrm{fused}}$ to be unsatisfiable, the solver exhibits a soundness bug.
For unsatisfiable seeds the construction is dual: the formulas are
disjoined and the constraint $z = x + y$ is added, so that the
fused formula remains unsatisfiable.
This can be seen as an example of metamorphic testing because
the expected satisfiability of the constructed formula is inherited from that of the seeds.

Metamorphic oracles such as EMI,
the semantics-preserving transformations employed by GraphicsFuzz,
and ideas like ternary logic partitioning and semantic fusion
have the feel of formal methods specifications because they
exploit semantic insights.
These insights allow fuzzing to go
beyond crash bugs, enabling deep functional correctness bugs to be found.

\paragraph*{Problems of scale?}

It would be an exaggeration to say that fuzzing is entirely free from problems of scale.
A fuzzer's success depends in large part on throughput, so fuzzers must be engineered for efficiency.
And the tool-scale problems of \Cref{sec:scale} have an analogue when the system under test takes a rich input format such as a programming language: the generator must then produce valid programs.
But by and large fuzzing is not burdened by problems of scale,
and importantly, it can be readily applied to very large software projects.

\section{Targeted Formal Methods Where Scalability Is Not Required}\label{sec:smallscale}

My change of focus from formal methods to fuzzing did not mean leaving formal methods behind entirely.
Instead, doubling down on the GPUVerify ethos of using formal methods to attack niche-but-important problems,
I gravitated towards problems where scale is not the blocker.
I now briefly describe two recent projects on the semantics of GPU programming languages where a small formal model,
developed in close collaboration with the engineers responsible for the language design, can have meaningful impact.

\begin{figure}
\centering
\includegraphics[width=\linewidth]{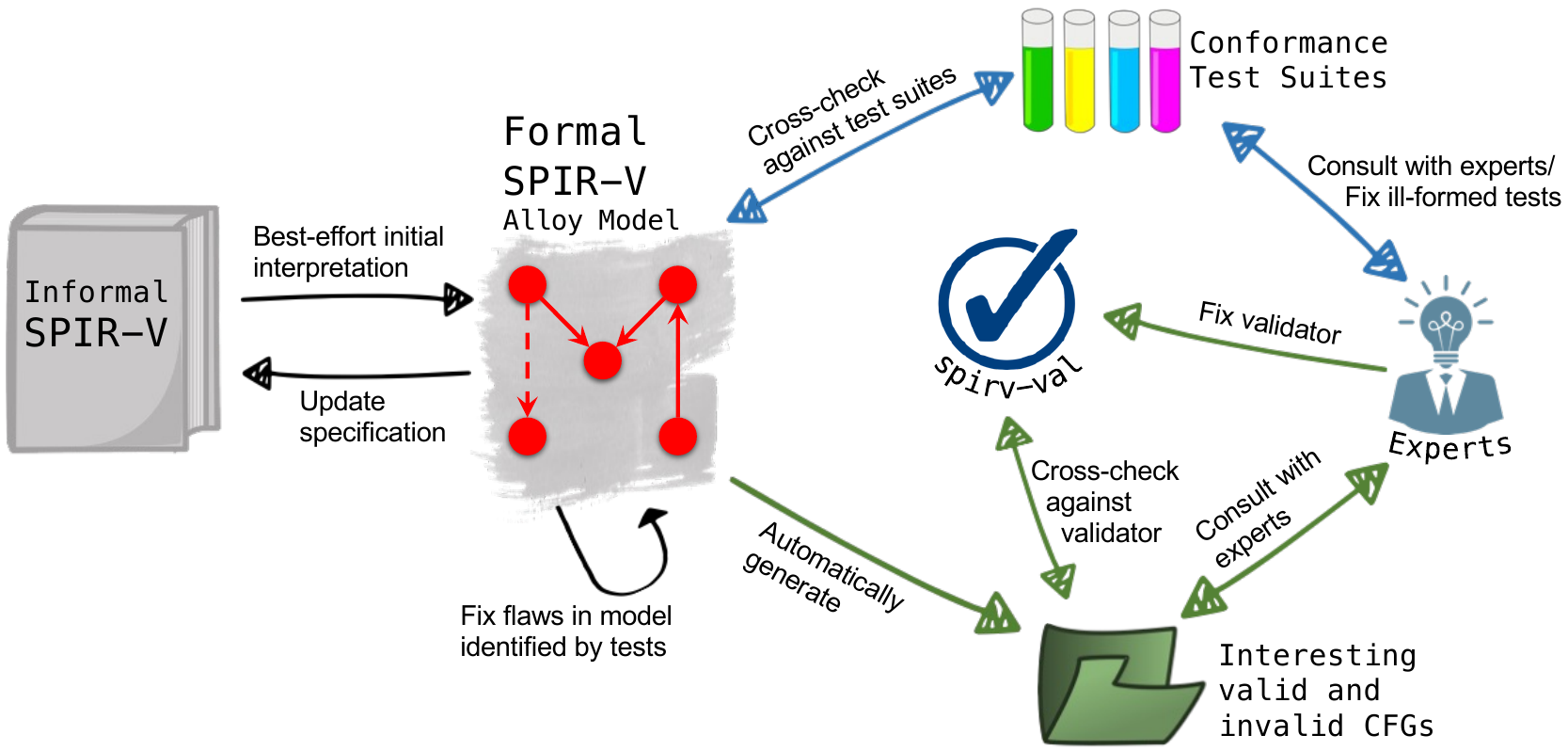}
\caption{The virtuous cycle of validating our Alloy model of SPIR-V against various sources of truth~\cite{DBLP:journals/pacmpl/KlimisCBNWD23}}
\label{fig:cross-validation}
\end{figure}

\paragraph*{Taking back control in SPIR-V}
A project I am particularly proud of concerned SPIR-V,
the industry-standard intermediate representation for GPU programming used by Vulkan and OpenCL~\cite{DBLP:journals/pacmpl/KlimisCBNWD23}.
SPIR-V is an unstructured, control-flow-graph-based language,
but it is equipped with annotations that allow high-level structured control flow, such as loops and selection constructs, to be expressed.
The rules governing these annotations had long been a source of confusion and frustration:
they were complex, ambiguous, and in places at odds with what their designers had intended.
In some cases their complexity had arisen from unsuccessful attempts to align them better with designer intentions.
My group at Imperial, working with SPIR-V experts at Google,
built a small formal model of SPIR-V control flow in the Alloy modelling language~\cite{DBLP:journals/cacm/Jackson19},
capable of determining whether a given control-flow graph is valid,
and of generating examples of valid and invalid control-flow graphs.
In the absence of a reliable specification, we started by writing a best-effort formal model,
faithfully modelling uncontroversial aspects of the SPIR-V specification, and including tentative fixes for parts of the specification we knew to be problematic.
We then iterated on our model by cross-checking it against several sources of truth:
the specification text, various SPIR-V test suites, the \texttt{spirv-val} validation tool,
and the expert judgement of our industrial collaborators.
Whenever the model disagreed with one of these sources,
the discrepancy revealed a problem in at least one of the model, the tests, the validator, or the specification itself, which we fixed.
This process led us to a new concept, \emph{structural dominance},
through which the control flow rules of SPIR-V can be stated accurately and far more simply.
Our reformulated rules have been adopted in the SPIR-V specification,
and along the way we fixed problems in the conformance tests and in \texttt{spirv-val}~\cite{DBLP:journals/pacmpl/KlimisCBNWD23}.
\Cref{fig:cross-validation} illustrates this virtuous cycle.

\paragraph*{Uniformity analysis in WGSL}
More recently, we have taken a similar approach to the WebGPU Shading Language (WGSL),
whose specification mandates a \emph{uniformity analysis} to reject programs that might suffer from barrier divergence.
We gave a formal semantics that defines barrier divergence for a core of WGSL,
reformulated the analysis as concise formal rules,
used the semantics to expose soundness and precision problems in the analysis as specified,
leading to changes to the specification that have all been accepted,
and mechanised the analysis in Alloy~\cite{DBLP:journals/pacmpl/Lee-JonesWD26}.

\medskip
Neither of these projects required heroic proof effort:
in both cases the models are relatively small,
and the Alloy Analyzer explores them automatically up to a given bound.
And neither of the projects offered particularly good scalability---due to its use of a SAT solver behind the scenes, we found that the Alloy Analyzer would quickly struggle to produce examples for large graphs with intricate constraints, and (as is so often the case with symbolic methods) performance could be rather unpredictable.
This turned out not to matter much.
Alloy's design rests on the \emph{small scope hypothesis}~\cite{jackson2012software}:
most defects in a model can be exhibited by a small instance,
so exhaustive exploration of all instances up to a modest size is a surprisingly effective way of finding them.
The same reasoning applies to defects in whatever the model describes,
and our experience with SPIR-V and WGSL bore this out: the soundness and precision problems we found in the SPIR-V and WGSL specifications all manifested on modest control-flow graphs and programs, and once the specification issues had been fixed we were able
to push the analysis to graphs and programs well beyond the size at which the last problem had appeared,
without finding any more.
While this kind of bounded result does not rule out the possibility of larger problematic examples,
it gives a more systematic version of the kind of confidence that testing provides.

\section{Formal Methods Meets Fuzzing}\label{sec:fmfuzzing}

Having a foot in both the formal methods and fuzzing camps, I am naturally interested in work that sits at the boundary between them,
where each can help the other.
Formal models can be a source of randomised test inputs for fuzzing, with strong oracles coming for free from the model; and formal methods tools are themselves large, complex pieces of software in which fuzzing can find bugs.

\paragraph*{Formal methods for fuzzing}
A bonus of building the Alloy models described in \Cref{sec:smallscale} is that the Alloy Analyzer can generate example instances of them, via a SAT solver.
In the SPIR-V work, this gave us a ready supply of valid control-flow graphs
that are arbitrary in shape, unconstrained by any particular programming style.
These were ideal seeds for fuzzing.
We exploited this via a simple-but-effective technique that we call \emph{control-flow graph fleshing}~\cite{DBLP:journals/pacmpl/KlimisCBNWD23}.
Given a control-flow graph skeleton generated by Alloy,
we choose a random path through it,
and flesh the skeleton out into an executable program that is forced to follow this path and that records the path it actually takes.
If the recorded path differs from the expected one, the compiler is at fault.
This gives every test a strong oracle by construction,
and fuzzing via fleshing found numerous previously unknown bugs in SPIR-V compilers and translators from several vendors.
Similarly, in the WGSL work we used Alloy's ability to generate examples to stress-test the uniformity analysis implemented in the Chromium web browser.
Here the oracle comes from differential testing between the model and Chromium: if they disagree about whether a program passes the analysis, one of them has a bug~\cite{DBLP:journals/pacmpl/Lee-JonesWD26}.

\paragraph*{Fuzzing for formal methods}
Formal methods tools are software,
and there is a strong case that formal methods tools should be considered critical software because a defect in a formal methods tool has the potential to undermine the guarantees that the tool claims to provide.\footnote{Although I am excited about working on testing formal methods tools such as analysers and verifiers, I am not fully sold on the dangers that bugs in these tools \emph{really} pose; this was the subject of an interesting debate at a recent Dagstuhl seminar~\cite[\S4.1]{DBLP:journals/dagstuhl-reports/ChristakisDRS25}.}
It therefore makes sense to use fuzzing to find bugs in formal methods tools!

An early example of this is the use of Csmith, designed to test C compilers,
to test the Frama-C static analysis platform~\cite{DBLP:conf/nfm/CuoqMPPRYY12}.
This led to fifty bugs being found and fixed in Frama-C,
and in a pleasing twist, Frama-C in turn exposed bugs in Csmith,
in the form of generated programs that exhibited undefined behaviour despite Csmith being designed to avoid it.
In a similar vein, various researchers (including my collaborators and me) have applied fuzzing to the verifier and compiler of Dafny, a verification-aware programming language~\cite{DBLP:conf/icst/DonaldsonSTU24,DBLP:conf/issta/IrfanPRRT22,DBLP:journals/pacmpl/0001024a}.
Bugs in a verifying compiler's code generator are particularly insidious:
if the compiler silently generates wrong code, the guarantees established by verification at the source level are lost,
and the bug evades both code review and formal verification.

In the hardware domain, we have used fuzzing to find bugs in commercial C-to-RTL formal equivalence checkers,
tools that are trusted to the point of being used to authorise hardware design sign-off~\cite{pardalos2025equifuzz}.
Our tool, Equifuzz, generates rich SystemC designs that are known, by construction, to be equivalent to trivial hardware designs.
An equivalence checker that disagrees has a \emph{completeness} bug: it fails to prove equivalence of designs that really are equivalent.
The counterexample associated with a failed proof often contains values indicating the result the equivalence checker (wrongly) believes the SystemC design should compute,
and it is often possible to exploit this by asking the equivalence checker to (wrongly) prove that the SystemC design indeed produces the given value,
thus flipping the completeness bug into a (more serious) \emph{soundness} bug.
Equifuzz found several soundness and completeness bugs in major commercial equivalence checkers,
which have since been confirmed and fixed, improving trust in these tools.

\section{Artificial Intelligence: Scalable Formal Methods At Last?}\label{sec:ai}

As I noted in the introduction, the frontier models of late 2025 crossed a threshold with respect to formal proof.
They have also become much more capable at constructing software, especially when used agentically.
I now return to the four problems of scale from \Cref{sec:scale}
and consider, for each, how much LLMs might change the picture.
It is important to note that these potential benefits of scale are only available to practitioners with
access to state-of-the-art models,
and that scale comes at high compute costs with associated environmental impact.
Making AI both accessible and sustainable is a challenge for the AI revolution as a whole, and one that those of us who stand to benefit from it have a responsibility to help address.

\paragraph*{Proof scale}
Recall that the central burden of interactive theorem proving is in constructing laborious, detailed proofs,
and that the central burden of auto-active verification is in finding supporting annotations,
such as loop invariants.
Both share the property that it is hard for a human to come up with the required artifact,
but easy for a machine to check whether a \emph{candidate} artifact fits the bill.
This plays to the strengths of LLMs:
their stochastic nature makes them well suited to proposing plausible proofs and invariants,
but there is no need to trust what they propose,
because a proof assistant kernel will reject an incorrect proof,
and a verification engine will reject an invariant that does not hold or is not inductive.
An LLM can therefore be used as a tireless, fallible generator of candidates,
for consideration by a rigorous, non-AI-based checker.
Hallucinations and other inaccuracies can, at worst, waste time and resources.
Critically, as long as the checker does its job, they cannot lead to an incorrect proof being accepted.

As a result of these advances, within the formal methods and programming languages communities it is becoming increasingly routine to formalise definitions, semantics and results in Lean or Rocq.
Just a few years ago, such formalisations would only usually be attempted by particular research groups with the requisite years of expertise.

One thing that a proof checker \emph{cannot} catch is a mistake in the \emph{specification}---the theorem statement and the definitions on which the statement depends.
Special care must therefore be taken before letting AI loose on both specifications and proofs,
and carefully reviewing and testing specifications (including via fuzzing) is more important than ever in an age of AI-generated proofs.

\paragraph*{Algorithmic scale}
Here I am less sure that AI changes the picture fundamentally.
The limits on automatic verification are limits of intractability and undecidability,
and a clever language model cannot make the state space of a system smaller!\footnote{One might object that proof search is intractable too, and yet AI has shown great promise.
My sense is that writing a proof is more like programming than like solving an NP-hard instance,
and that this is why LLMs are good at it.}
That said, AI has the potential to help around the edges of algorithmic scale.
For example, AI could absorb some of the brittleness associated with SMT-based verification described in \Cref{sec:scale}:
an agent could watch for cases where previously successful verification breaks
after a tool or configuration update, and then iterate the process of
(for example) updating solver options, modifying encodings, adjusting invariants and
fine-tuning E-matching triggers, until verification succeeds again.
This is exactly the kind of tedious, well-specified task that AI is good at.
Furthermore, AI may serve to accelerate research on verification algorithms themselves,
because LLMs make it cheap to prototype an idea and, increasingly, to formally verify that the idea is sound.

\paragraph*{Tool scale}
The opportunity afforded by AI here is simple:
LLMs make it much cheaper to create software, or at least initial prototypes of software.
Whether they will transform the creation of high-quality, long-lived software remains to be seen.
But proof-of-concept tooling of the kind needed to demonstrate a formal method in action---extending a front end to handle the esoteric language features that one particular codebase relies on,
bridging a solver-based method to a different solver,
wiring a verifier into a build system---is precisely the sort of engineering work that agentic coding tools are now very good at.
In \Cref{sec:scale} I noted that formal methods researchers are often unwilling or unable to do the tooling work needed to apply a technique to real code;
in the age of LLMs that excuse now seems rather weak.

\paragraph*{Team scale}
There is potential for AI to help on both sides of the team-scale problem:
accelerating the acquisition of sufficient expertise in a formal method,
and making it easier to distribute formal methods work across a team.

On the first point, I have personally found LLMs to be enormously useful in learning Lean,
quickly allowing me to get to the point where I can do meaningful work in it.
It also seems clear that ``on-tap'' access to AI assistance lowers the level of expertise one needs to use a formal method productively.\footnote{As an aside, I do \emph{not} expect AI to do much for the acquisition of deep expertise;
digesting a complex formalism properly takes time,
and learning often comes from being lost for a while.
My hunch is that the ability to reach for an LLM every time one gets stuck
will accelerate short-term progress to the detriment of long-term understanding.
The likely outcome is fewer people with deep expertise and many more people who know enough to be effective.}

Second, there is growing evidence that LLMs are good at proof repair~\cite{DBLP:conf/sigsoft/FirstRRB23,DBLP:journals/corr/abs-2602-02990}, which is the core of proof maintenance.
If proofs can be kept in step with code largely automatically,
the agile style of development that I argued in \Cref{sec:scale} is at odds with formal methods
may become practical for formal methods too.
DeMillo, Lipton and Perlis's argument that verifying a modified program is as hard as verifying the original would then lose much of its force.

\paragraph*{The Keystone project}
With collaborators at Imperial College London and EPFL,
I recently embarked on Keystone~\cite{aria2026keystone},
a project which seeks to investigate the impact that AI can have on formal methods at scale.
Keystone is funded through a 2026 call from ARIA's Safeguarded AI programme on using AI-enabled formal methods for cybersecurity~\cite{aria2026cybersecurity}.
The premise of the call aligns closely with the argument of this paper:
it observes that formal methods have historically been too slow and expertise-intensive to apply at scale,
that AI is strong at exactly the kind of search-intensive work that traditionally made formal methods hard to apply,
and that ``speed from AI does not require trust in AI''~\cite{aria2026cybersecurity}.
Funded teams are asked to build production-grade, security-critical components with machine-checked proofs of their key properties, using AI as a central means of reaching a level of ambition that would otherwise be out of reach.

Our focus in Keystone is on creating LLM inference serving infrastructure with strong, machine-checked correctness guarantees.
Inference serving software is now a form of critical infrastructure:
the routine dependence of businesses, governments and members of the public on LLMs now means that
denial-of-service vulnerabilities or isolation failures between tenants whose queries are being served via the same GPU
have the potential to be very serious.

An appealing feature of inference as a verification target is that, unusually for a large software system, it has a clean specification:
a transformer computes the next token as a well-defined mathematical function of the model's weights and the embeddings of the preceding tokens.
Showing that a high-performance inference implementation---with optimisations such as paged attention, prefix caching and continuous batching---refines that function is a challenging but well-defined task, and I am excited about the progress we can make thanks to the assistance that AI can provide.

\section*{Acknowledgements}

I am grateful to Amber Gorzynski, James Lee-Jones, Manuel Rigger and John Wickerson for their feedback on drafts of this paper.

The work that I summarise here was done in collaboration with many Masters and PhD students at Imperial, postdoctoral and faculty colleagues at Imperial,
and academic and industrial collaborators internationally. I am grateful to all of them.

The work was supported by numerous grants over the years,
including the EU FP7 STREP project CARP (project number 287767),
an EPSRC Early Career Fellowship (EP/N026314/1),
the EPSRC IRIS Programme Grant (EP/R006865/1),
two Amazon Research Awards,
gift funding from Google,
and a grant from the Advanced Research and Invention Agency (ARIA).

\bibliography{references}

@inproceedings{DBLP:conf/tacas/MouraB08,
  author       = {Leonardo Mendon{\c{c}}a de Moura and
                  Nikolaj S. Bj{\o}rner},
  editor       = {C. R. Ramakrishnan and
                  Jakob Rehof},
  title        = {{Z3:} An Efficient {SMT} Solver},
  booktitle    = {Tools and Algorithms for the Construction and Analysis of Systems,
                  14th International Conference, {TACAS} 2008, Held as Part of the Joint
                  European Conferences on Theory and Practice of Software, {ETAPS} 2008,
                  Budapest, Hungary, March 29-April 6, 2008. Proceedings},
  series       = {Lecture Notes in Computer Science},
  volume       = {4963},
  pages        = {337--340},
  publisher    = {Springer},
  year         = {2008},
  doi          = {10.1007/978-3-540-78800-3_24},
}

@inproceedings{DBLP:conf/cav/BarrettCDHJKRT11,
  author       = {Clark W. Barrett and
                  Christopher L. Conway and
                  Morgan Deters and
                  Liana Hadarean and
                  Dejan Jovanovic and
                  Tim King and
                  Andrew Reynolds and
                  Cesare Tinelli},
  editor       = {Ganesh Gopalakrishnan and
                  Shaz Qadeer},
  title        = {{CVC4}},
  booktitle    = {Computer Aided Verification - 23rd International Conference, {CAV}
                  2011, Snowbird, UT, USA, July 14-20, 2011. Proceedings},
  series       = {Lecture Notes in Computer Science},
  volume       = {6806},
  pages        = {171--177},
  publisher    = {Springer},
  year         = {2011},
  doi          = {10.1007/978-3-642-22110-1_14},
}

@misc{aria2026cybersecurity,
  author       = {{Advanced Research and Invention Agency}},
  title        = {Safeguarded {AI}: Cybersecurity -- {Call for Proposals}},
  year         = {2026},
  month        = may,
  howpublished = {\url{https://aria.org.uk/media/1padxpaf/safeguarded-ai_cybersecurity_solicitation.pdf}},
  note         = {Version 1.0, 20 May 2026. Accessed 15 October 2026}
}

@article{DBLP:journals/corr/abs-2602-02990,
  author       = {Evan Wang and
                  Simon Chess and
                  Daniel Lee and
                  Siyuan Ge and
                  Ajit Mallavarapu and
                  Vasily Ilin},
  title        = {Learning to Repair Lean Proofs from Compiler Feedback},
  journal      = {CoRR},
  volume       = {abs/2602.02990},
  year         = {2026},
  doi          = {10.48550/ARXIV.2602.02990},
  eprinttype   = {arXiv},
  eprint       = {2602.02990},
}

@inproceedings{DBLP:conf/sigsoft/FirstRRB23,
  author       = {Emily First and
                  Markus N. Rabe and
                  Talia Ringer and
                  Yuriy Brun},
  editor       = {Satish Chandra and
                  Kelly Blincoe and
                  Paolo Tonella},
  title        = {Baldur: Whole-Proof Generation and Repair with Large Language Models},
  booktitle    = {Proceedings of the 31st {ACM} Joint European Software Engineering
                  Conference and Symposium on the Foundations of Software Engineering,
                  {ESEC/FSE} 2023, San Francisco, CA, USA, December 3-9, 2023},
  pages        = {1229--1241},
  publisher    = {{ACM}},
  year         = {2023},
  url          = {https://doi.org/10.1145/3611643.3616243},
  doi          = {10.1145/3611643.3616243},
  bibsource    = {dblp computer science bibliography, https://dblp.org}
}

@inproceedings{DBLP:conf/cgo/LattnerA04,
  author       = {Chris Lattner and
                  Vikram S. Adve},
  title        = {{LLVM:} {A} Compilation Framework for Lifelong Program Analysis {\&}
                  Transformation},
  booktitle    = {2nd {IEEE} / {ACM} International Symposium on Code Generation and
                  Optimization {(CGO} 2004), 20-24 March 2004, San Jose, CA, {USA}},
  pages        = {75--88},
  publisher    = {{IEEE} Computer Society},
  year         = {2004},
  doi          = {10.1109/CGO.2004.1281665},
}

@inproceedings{DBLP:conf/cav/BardsleyBCCDDKLQ14,
  author       = {Ethel Bardsley and
                  Adam Betts and
                  Nathan Chong and
                  Peter Collingbourne and
                  Pantazis Deligiannis and
                  Alastair F. Donaldson and
                  Jeroen Ketema and
                  Daniel Liew and
                  Shaz Qadeer},
  editor       = {Armin Biere and
                  Roderick Bloem},
  title        = {Engineering a Static Verification Tool for {GPU} Kernels},
  booktitle    = {Computer Aided Verification - 26th International Conference, {CAV}
                  2014, Held as Part of the Vienna Summer of Logic, {VSL} 2014, Vienna,
                  Austria, July 18-22, 2014. Proceedings},
  series       = {Lecture Notes in Computer Science},
  volume       = {8559},
  pages        = {226--242},
  publisher    = {Springer},
  year         = {2014},
  doi          = {10.1007/978-3-319-08867-9_15},
}

@inproceedings{DBLP:conf/cav/BeyerHT07,
  author       = {Dirk Beyer and
                  Thomas A. Henzinger and
                  Gr{\'{e}}gory Th{\'{e}}oduloz},
  editor       = {Werner Damm and
                  Holger Hermanns},
  title        = {Configurable Software Verification: Concretizing the Convergence of
                  Model Checking and Program Analysis},
  booktitle    = {Computer Aided Verification, 19th International Conference, {CAV}
                  2007, Berlin, Germany, July 3-7, 2007, Proceedings},
  series       = {Lecture Notes in Computer Science},
  volume       = {4590},
  pages        = {504--518},
  publisher    = {Springer},
  year         = {2007},
  doi          = {10.1007/978-3-540-73368-3_51},
}

@inproceedings{DBLP:conf/cav/BeyerK11,
  author       = {Dirk Beyer and
                  M. Erkan Keremoglu},
  editor       = {Ganesh Gopalakrishnan and
                  Shaz Qadeer},
  title        = {{CPAchecker}: {A} Tool for Configurable Software Verification},
  booktitle    = {Computer Aided Verification - 23rd International Conference, {CAV}
                  2011, Snowbird, UT, USA, July 14-20, 2011. Proceedings},
  series       = {Lecture Notes in Computer Science},
  volume       = {6806},
  pages        = {184--190},
  publisher    = {Springer},
  year         = {2011},
  doi          = {10.1007/978-3-642-22110-1_16},
}

@inproceedings{DBLP:conf/iwpc/JuhosovaZC25,
  author       = {S{\'{a}}ra Juho{\v{s}}ov{\'{a}} and
                  Andy Zaidman and
                  Jesper Cockx},
  title        = {Pinpointing the Learning Obstacles of an Interactive Theorem Prover},
  booktitle    = {33rd {IEEE/ACM} International Conference on Program Comprehension,
                  ICPC@ICSE 2025, Ottawa, ON, Canada, April 27-28, 2025},
  pages        = {159--170},
  publisher    = {{IEEE}},
  year         = {2025},
  doi          = {10.1109/ICPC66645.2025.00024},
}

@article{DBLP:journals/ftpl/RingerPSGT19,
  author       = {Talia Ringer and
                  Karl Palmskog and
                  Ilya Sergey and
                  Milos Gligoric and
                  Zachary Tatlock},
  title        = {{QED} at Large: {A} Survey of Engineering of Formally Verified Software},
  journal      = {Found. Trends Program. Lang.},
  volume       = {5},
  number       = {2-3},
  pages        = {102--281},
  year         = {2019},
  doi          = {10.1561/2500000045},
}

@article{DBLP:journals/dagstuhl-reports/ChristakisDRS25,
  author       = {Maria Christakis and
                  Alastair F. Donaldson and
                  John Regehr and
                  Thodoris Sotiropoulos},
  title        = {Testing Program Analyzers and Verifiers ({Dagstuhl} {Seminar} 25242)},
  journal      = {Dagstuhl Reports},
  volume       = {15},
  number       = {6},
  pages        = {69--83},
  year         = {2025},
  doi          = {10.4230/DAGREP.15.6.69},
}

@article{DBLP:journals/eceasst/Donaldson14,
  author       = {Alastair F. Donaldson},
  title        = {The {GPUVerify} Method: a Tutorial Overview},
  journal      = {Electron. Commun. Eur. Assoc. Softw. Sci. Technol.},
  volume       = {70},
  year         = {2014},
  doi          = {10.14279/TUJ.ECEASST.70.986},
}

@article{DBLP:journals/software/Hall90,
  author       = {Anthony Hall},
  title        = {Seven Myths of Formal Methods},
  journal      = {{IEEE} Softw.},
  volume       = {7},
  number       = {5},
  pages        = {11--19},
  year         = {1990},
  doi          = {10.1109/52.57887},
}

@book{jackson2012software,
  author    = {Daniel Jackson},
  title     = {Software Abstractions: Logic, Language, and Analysis},
  edition   = {Revised},
  publisher = {{MIT} Press},
  address   = {Cambridge, MA},
  year      = {2012},
  isbn      = {978-0-262-01715-2}
}

@inproceedings{DBLP:conf/icse/AndronickJKKSZZ12,
  author       = {June Andronick and
                  D. Ross Jeffery and
                  Gerwin Klein and
                  Rafal Kolanski and
                  Mark Staples and
                  He Zhang and
                  Liming Zhu},
  editor       = {Martin Glinz and
                  Gail C. Murphy and
                  Mauro Pezz{\`{e}}},
  title        = {Large-scale formal verification in practice: {A} process perspective},
  booktitle    = {34th International Conference on Software Engineering, {ICSE} 2012,
                  June 2-9, 2012, Zurich, Switzerland},
  pages        = {1002--1011},
  publisher    = {{IEEE} Computer Society},
  year         = {2012},
  doi          = {10.1109/ICSE.2012.6227120},
}

@inproceedings{DBLP:conf/aisc/BourkeDKK12,
  author       = {Timothy Bourke and
                  Matthias Daum and
                  Gerwin Klein and
                  Rafal Kolanski},
  editor       = {Johan Jeuring and
                  John A. Campbell and
                  Jacques Carette and
                  Gabriel Dos Reis and
                  Petr Sojka and
                  Makarius Wenzel and
                  Volker Sorge},
  title        = {Challenges and Experiences in Managing Large-Scale Proofs},
  booktitle    = {Intelligent Computer Mathematics - 11th International Conference,
                  {AISC} 2012, 19th Symposium, Calculemus 2012, 5th International Workshop,
                  {DML} 2012, 11th International Conference, {MKM} 2012, Systems and
                  Projects, Held as Part of {CICM} 2012, Bremen, Germany, July 8-13,
                  2012. Proceedings},
  series       = {Lecture Notes in Computer Science},
  volume       = {7362},
  pages        = {32--48},
  publisher    = {Springer},
  year         = {2012},
  doi          = {10.1007/978-3-642-31374-5_3},
}

@article{DBLP:journals/cj/Weyuker82,
  author       = {Elaine J. Weyuker},
  title        = {On Testing Non-Testable Programs},
  journal      = {Comput. J.},
  volume       = {25},
  number       = {4},
  pages        = {465--470},
  year         = {1982},
  doi          = {10.1093/COMJNL/25.4.465},
}

@inproceedings{DBLP:conf/fm/FlanaganL01,
  author       = {Cormac Flanagan and
                  K. Rustan M. Leino},
  editor       = {Jos{\'{e}} Nuno Oliveira and
                  Pamela Zave},
  title        = {Houdini, an Annotation Assistant for {ESC/Java}},
  booktitle    = {{FME} 2001: Formal Methods for Increasing Software Productivity, International
                  Symposium of Formal Methods Europe, Berlin, Germany, March 12-16,
                  2001, Proceedings},
  series       = {Lecture Notes in Computer Science},
  volume       = {2021},
  pages        = {500--517},
  publisher    = {Springer},
  year         = {2001},
  doi          = {10.1007/3-540-45251-6_29},
}

@inproceedings{DBLP:conf/tacas/ClarkeKL04,
  author       = {Edmund M. Clarke and
                  Daniel Kroening and
                  Flavio Lerda},
  editor       = {Kurt Jensen and
                  Andreas Podelski},
  title        = {A Tool for Checking {ANSI-C} Programs},
  booktitle    = {Tools and Algorithms for the Construction and Analysis of Systems,
                  10th International Conference, {TACAS} 2004, Held as Part of the Joint
                  European Conferences on Theory and Practice of Software, {ETAPS} 2004,
                  Barcelona, Spain, March 29 - April 2, 2004, Proceedings},
  series       = {Lecture Notes in Computer Science},
  volume       = {2988},
  pages        = {168--176},
  publisher    = {Springer},
  year         = {2004},
  doi          = {10.1007/978-3-540-24730-2_15},
}

@inproceedings{DBLP:conf/fmco/BarnettCDJL05,
  author       = {Michael Barnett and
                  Bor{-}Yuh Evan Chang and
                  Robert DeLine and
                  Bart Jacobs and
                  K. Rustan M. Leino},
  editor       = {Frank S. de Boer and
                  Marcello M. Bonsangue and
                  Susanne Graf and
                  Willem P. de Roever},
  title        = {Boogie: {A} Modular Reusable Verifier for Object-Oriented Programs},
  booktitle    = {Formal Methods for Components and Objects, 4th International Symposium,
                  {FMCO} 2005, Amsterdam, The Netherlands, November 1-4, 2005, Revised
                  Lectures},
  series       = {Lecture Notes in Computer Science},
  volume       = {4111},
  pages        = {364--387},
  publisher    = {Springer},
  year         = {2005},
  doi          = {10.1007/11804192_17},
}

@inproceedings{DBLP:conf/oopsla/ChongDKKQ13,
  author       = {Nathan Chong and
                  Alastair F. Donaldson and
                  Paul H. J. Kelly and
                  Jeroen Ketema and
                  Shaz Qadeer},
  editor       = {Antony L. Hosking and
                  Patrick Th. Eugster and
                  Cristina V. Lopes},
  title        = {Barrier invariants: a shared state abstraction for the analysis of
                  data-dependent {GPU} kernels},
  booktitle    = {Proceedings of the 2013 {ACM} {SIGPLAN} International Conference on
                  Object Oriented Programming Systems Languages {\&} Applications,
                  {OOPSLA} 2013, part of {SPLASH} 2013, Indianapolis, IN, USA, October
                  26-31, 2013},
  pages        = {605--622},
  publisher    = {{ACM}},
  year         = {2013},
  doi          = {10.1145/2509136.2509517},
}

@article{DBLP:journals/cacm/Jackson19,
  author       = {Daniel Jackson},
  title        = {Alloy: a language and tool for exploring software designs},
  journal      = {Commun. {ACM}},
  volume       = {62},
  number       = {9},
  pages        = {66--76},
  year         = {2019},
  doi          = {10.1145/3338843},
}

@inproceedings{pardalos2025equifuzz,
  author       = {Michalis Pardalos and
                  Alastair F. Donaldson and
                  Emiliano Morini and
                  Laura Pozzi and
                  John Wickerson},
  title        = {Who checks the checkers? {Automatically} finding bugs in {C}-to-{RTL} formal equivalence checkers},
  booktitle    = {DVCon Europe},
  year         = {2024},
  doi          = {10.30420/566438006},
}

@article{DBLP:journals/tc/Bryant86,
  author       = {Randal E. Bryant},
  title        = {Graph-Based Algorithms for {Boolean} Function Manipulation},
  journal      = {{IEEE} Trans. Computers},
  volume       = {35},
  number       = {8},
  pages        = {677--691},
  year         = {1986},
  doi          = {10.1109/TC.1986.1676819},
}

@misc{leino2010autoactive,
  author       = {K. Rustan M. Leino and Micha{\l} Moskal},
  title        = {Usable Auto-Active Verification},
  month        = nov,
  year         = {2010},
  url         = {https://leino.science/papers/krml212.pdf}
}

@inproceedings{DBLP:conf/fmcad/ZhouBTLHP23,
  author       = {Yi Zhou and
                  Jay Bosamiya and
                  Yoshiki Takashima and
                  Jessica Li and
                  Marijn Heule and
                  Bryan Parno},
  editor       = {Alexander Nadel and
                  Kristin Yvonne Rozier},
  title        = {Mariposa: Measuring {SMT} Instability in Automated Program Verification},
  booktitle    = {Formal Methods in Computer-Aided Design, {FMCAD} 2023, Ames, IA, USA,
                  October 24-27, 2023},
  pages        = {178--188},
  publisher    = {{IEEE}},
  year         = {2023},
  doi          = {10.34727/2023/ISBN.978-3-85448-060-0_26},
}

@inproceedings{DBLP:conf/popl/SwamyHKRDFBFSKZ16,
  author       = {Nikhil Swamy and
                  Catalin Hritcu and
                  Chantal Keller and
                  Aseem Rastogi and
                  Antoine Delignat{-}Lavaud and
                  Simon Forest and
                  Karthikeyan Bhargavan and
                  C{\'{e}}dric Fournet and
                  Pierre{-}Yves Strub and
                  Markulf Kohlweiss and
                  Jean Karim Zinzindohoue and
                  Santiago Zanella{-}B{\'{e}}guelin},
  editor       = {Rastislav Bod{\'{\i}}k and
                  Rupak Majumdar},
  title        = {Dependent types and multi-monadic effects in {F$^\star$}},
  booktitle    = {Proceedings of the 43rd Annual {ACM} {SIGPLAN-SIGACT} Symposium on
                  Principles of Programming Languages, {POPL} 2016, St. Petersburg,
                  FL, USA, January 20 - 22, 2016},
  pages        = {256--270},
  publisher    = {{ACM}},
  year         = {2016},
  doi          = {10.1145/2837614.2837655},
}

@inproceedings{DBLP:conf/lpar/Leino10,
  author       = {K. Rustan M. Leino},
  editor       = {Edmund M. Clarke and
                  Andrei Voronkov},
  title        = {Dafny: An Automatic Program Verifier for Functional Correctness},
  booktitle    = {Logic for Programming, Artificial Intelligence, and Reasoning - 16th
                  International Conference, LPAR-16, Dakar, Senegal, April 25-May 1,
                  2010, Revised Selected Papers},
  series       = {Lecture Notes in Computer Science},
  volume       = {6355},
  pages        = {348--370},
  publisher    = {Springer},
  year         = {2010},
  doi          = {10.1007/978-3-642-17511-4_20},
}

@inproceedings{DBLP:conf/esop/FilliatreP13,
  author       = {Jean{-}Christophe Filli{\^{a}}tre and
                  Andrei Paskevich},
  editor       = {Matthias Felleisen and
                  Philippa Gardner},
  title        = {Why3 -- Where Programs Meet Provers},
  booktitle    = {Programming Languages and Systems - 22nd European Symposium on Programming,
                  {ESOP} 2013, Held as Part of the European Joint Conferences on Theory
                  and Practice of Software, {ETAPS} 2013, Rome, Italy, March 16-24,
                  2013. Proceedings},
  series       = {Lecture Notes in Computer Science},
  volume       = {7792},
  pages        = {125--128},
  publisher    = {Springer},
  year         = {2013},
  doi          = {10.1007/978-3-642-37036-6_8},
}

@inproceedings{DBLP:conf/vmcai/0001SS16,
  author       = {Peter M{\"{u}}ller and
                  Malte Schwerhoff and
                  Alexander J. Summers},
  editor       = {Barbara Jobstmann and
                  K. Rustan M. Leino},
  title        = {Viper: {A} Verification Infrastructure for Permission-Based Reasoning},
  booktitle    = {Verification, Model Checking, and Abstract Interpretation - 17th International
                  Conference, {VMCAI} 2016, St. Petersburg, FL, USA, January 17-19,
                  2016. Proceedings},
  series       = {Lecture Notes in Computer Science},
  volume       = {9583},
  pages        = {41--62},
  publisher    = {Springer},
  year         = {2016},
  doi          = {10.1007/978-3-662-49122-5_2},
}

@inproceedings{DBLP:conf/cade/MouraB07,
  author       = {Leonardo Mendon{\c{c}}a de Moura and
                  Nikolaj S. Bj{\o}rner},
  editor       = {Frank Pfenning},
  title        = {Efficient {E}-Matching for {SMT} Solvers},
  booktitle    = {Automated Deduction - CADE-21, 21st International Conference on Automated
                  Deduction, Bremen, Germany, July 17-20, 2007, Proceedings},
  series       = {Lecture Notes in Computer Science},
  volume       = {4603},
  pages        = {183--198},
  publisher    = {Springer},
  year         = {2007},
  doi          = {10.1007/978-3-540-73595-3_13},
}

@inproceedings{DBLP:conf/icse/ChakarovGHHHJJL25,
  author       = {Aleks Chakarov and
                  Jaco Geldenhuys and
                  Matthew Heck and
                  Michael Hicks and
                  Sam Huang and
                  Georges{-}Axel Jaloyan and
                  Anjali Joshi and
                  K. Rustan M. Leino and
                  Mikael Mayer and
                  Sean McLaughlin and
                  Akhilesh Mritunjai and
                  Cl{\'{e}}ment Pit{-}Claudel and
                  Sorawee Porncharoenwase and
                  Florian Rabe and
                  Marianna Rapoport and
                  Giles Reger and
                  Cody Roux and
                  Neha Rungta and
                  Robin Salkeld and
                  Matthias Schlaipfer and
                  Daniel Schoepe and
                  Johanna Schwartzentruber and
                  Serdar Tasiran and
                  Aaron Tomb and
                  Emina Torlak and
                  Jean{-}Baptiste Tristan and
                  Lucas G. Wagner and
                  Michael W. Whalen and
                  Remy Willems and
                  Tongtong Xiang and
                  Taejoon Byun and
                  Joshua M. Cohen and
                  Ruijie Fang and
                  Junyoung Jang and
                  Jakob Rath and
                  Hira Taqdees Syeda and
                  Dominik Wagner and
                  Yongwei Yuan},
  title        = {Formally Verified Cloud-Scale Authorization},
  booktitle    = {47th {IEEE/ACM} International Conference on Software Engineering,
                  {ICSE} 2025, Ottawa, ON, Canada, April 26 - May 6, 2025},
  pages        = {2508--2521},
  publisher    = {{IEEE}},
  year         = {2025},
  doi          = {10.1109/ICSE55347.2025.00166},
}

@misc{tomb2023brittleness,
  author       = {Aaron Tomb and Jean-Baptiste Tristan},
  title        = {Avoiding Verification Brittleness in {Dafny}},
  howpublished = {Dafny Blog, \url{https://dafny.org/blog/2023/12/01/avoiding-verification-brittleness/}},
  year         = {2023},
  month        = dec,
  note         = {Accessed 13 October 2026}
}

@misc{salkeld2022cost,
  author       = {Robin Salkeld},
  title        = {Proof, but at What Cost?},
  howpublished = {Talk at the 22nd Annual High Confidence Software and Systems Conference ({HCSS} 2022)},
  year         = {2022},
  month        = may,
  note         = {Abstract and recording available at \url{https://sos-vo.org/node/83866}}
}

@article{DBLP:journals/cacm/SadowskiAEMJ18,
  author       = {Caitlin Sadowski and
                  Edward Aftandilian and
                  Alex Eagle and
                  Liam Miller{-}Cushon and
                  Ciera Jaspan},
  title        = {Lessons from building static analysis tools at {Google}},
  journal      = {Commun. {ACM}},
  volume       = {61},
  number       = {4},
  pages        = {58--66},
  year         = {2018},
  doi          = {10.1145/3188720},
}

@article{DBLP:journals/cacm/DistefanoFLO19,
  author       = {Dino Distefano and
                  Manuel F{\"{a}}hndrich and
                  Francesco Logozzo and
                  Peter W. O'Hearn},
  title        = {Scaling static analyses at {Facebook}},
  journal      = {Commun. {ACM}},
  volume       = {62},
  number       = {8},
  pages        = {62--70},
  year         = {2019},
  doi          = {10.1145/3338112},
}

@article{DBLP:journals/cacm/BesseyBCCFHHKME10,
  author       = {Al Bessey and
                  Ken Block and
                  Benjamin Chelf and
                  Andy Chou and
                  Bryan Fulton and
                  Seth Hallem and
                  Charles{-}Henri Gros and
                  Asya Kamsky and
                  Scott McPeak and
                  Dawson R. Engler},
  title        = {A few billion lines of code later: using static analysis to find bugs
                  in the real world},
  journal      = {Commun. {ACM}},
  volume       = {53},
  number       = {2},
  pages        = {66--75},
  year         = {2010},
  doi          = {10.1145/1646353.1646374},
}

@inproceedings{DBLP:conf/cav/ClarkeGJLV00,
  author       = {Edmund M. Clarke and
                  Orna Grumberg and
                  Somesh Jha and
                  Yuan Lu and
                  Helmut Veith},
  editor       = {E. Allen Emerson and
                  A. Prasad Sistla},
  title        = {Counterexample-Guided Abstraction Refinement},
  booktitle    = {Computer Aided Verification, 12th International Conference, {CAV}
                  2000, Chicago, IL, USA, July 15-19, 2000, Proceedings},
  series       = {Lecture Notes in Computer Science},
  volume       = {1855},
  pages        = {154--169},
  publisher    = {Springer},
  year         = {2000},
  doi          = {10.1007/10722167_15},
}

@article{DBLP:journals/tse/CordeiroFM12,
  author       = {Lucas C. Cordeiro and
                  Bernd Fischer and
                  Jo{\~{a}}o Marques{-}Silva},
  title        = {{SMT}-Based Bounded Model Checking for Embedded {ANSI-C} Software},
  journal      = {{IEEE} Trans. Software Eng.},
  volume       = {38},
  number       = {4},
  pages        = {957--974},
  year         = {2012},
  doi          = {10.1109/TSE.2011.59},
}

@article{DBLP:journals/sttt/ArmandoMP09,
  author       = {Alessandro Armando and
                  Jacopo Mantovani and
                  Lorenzo Platania},
  title        = {Bounded model checking of software using {SMT} solvers instead of
                  {SAT} solvers},
  journal      = {Int. J. Softw. Tools Technol. Transf.},
  volume       = {11},
  number       = {1},
  pages        = {69--83},
  year         = {2009},
  doi          = {10.1007/S10009-008-0091-0},
}

@inproceedings{DBLP:conf/tacas/BiereCCZ99,
  author       = {Armin Biere and
                  Alessandro Cimatti and
                  Edmund M. Clarke and
                  Yunshan Zhu},
  editor       = {Rance Cleaveland},
  title        = {Symbolic Model Checking without {BDDs}},
  booktitle    = {Tools and Algorithms for Construction and Analysis of Systems, 5th
                  International Conference, {TACAS} '99, Held as Part of the European
                  Joint Conferences on the Theory and Practice of Software, ETAPS'99,
                  Amsterdam, The Netherlands, March 22-28, 1999, Proceedings},
  series       = {Lecture Notes in Computer Science},
  volume       = {1579},
  pages        = {193--207},
  publisher    = {Springer},
  year         = {1999},
  doi          = {10.1007/3-540-49059-0_14},
}

@article{DBLP:journals/iandc/BurchCMDH92,
  author       = {Jerry R. Burch and
                  Edmund M. Clarke and
                  Kenneth L. McMillan and
                  David L. Dill and
                  L. J. Hwang},
  title        = {Symbolic Model Checking: $10^{20}$ States and Beyond},
  journal      = {Inf. Comput.},
  volume       = {98},
  number       = {2},
  pages        = {142--170},
  year         = {1992},
  doi          = {10.1016/0890-5401(92)90017-A},
}

@book{DBLP:books/sp/Godefroid96,
  author       = {Patrice Godefroid},
  title        = {Partial-Order Methods for the Verification of Concurrent Systems -
                  An Approach to the State-Explosion Problem},
  series       = {Lecture Notes in Computer Science},
  volume       = {1032},
  publisher    = {Springer},
  year         = {1996},
  doi          = {10.1007/3-540-60761-7},
  isbn         = {3-540-60761-7},
}

@article{DBLP:journals/toplas/ClarkeGL94,
  author       = {Edmund M. Clarke and
                  Orna Grumberg and
                  David E. Long},
  title        = {Model Checking and Abstraction},
  journal      = {{ACM} Trans. Program. Lang. Syst.},
  volume       = {16},
  number       = {5},
  pages        = {1512--1542},
  year         = {1994},
  doi          = {10.1145/186025.186051},
}

@inproceedings{DBLP:conf/popl/HenzingerJMS02,
  author       = {Thomas A. Henzinger and
                  Ranjit Jhala and
                  Rupak Majumdar and
                  Gr{\'{e}}goire Sutre},
  editor       = {John Launchbury and
                  John C. Mitchell},
  title        = {Lazy abstraction},
  booktitle    = {Conference Record of {POPL} 2002: The 29th {SIGPLAN-SIGACT} Symposium
                  on Principles of Programming Languages, Portland, OR, USA, January
                  16-18, 2002},
  pages        = {58--70},
  publisher    = {{ACM}},
  year         = {2002},
  doi          = {10.1145/503272.503279},
}

@inproceedings{DBLP:conf/pldi/BallMMR01,
  author       = {Thomas Ball and
                  Rupak Majumdar and
                  Todd D. Millstein and
                  Sriram K. Rajamani},
  editor       = {Michael Burke and
                  Mary Lou Soffa},
  title        = {Automatic Predicate Abstraction of {C} Programs},
  booktitle    = {Proceedings of the 2001 {ACM} {SIGPLAN} Conference on Programming
                  Language Design and Implementation (PLDI), Snowbird, Utah, USA, June
                  20-22, 2001},
  pages        = {203--213},
  publisher    = {{ACM}},
  year         = {2001},
  doi          = {10.1145/378795.378846},
}

@article{DBLP:journals/fmsd/EmersonS96,
  author       = {E. Allen Emerson and
                  A. Prasad Sistla},
  title        = {Symmetry and Model Checking},
  journal      = {Formal Methods Syst. Des.},
  volume       = {9},
  number       = {1/2},
  pages        = {105--131},
  year         = {1996},
  doi          = {10.1007/BF00625970},
}

@article{DBLP:journals/fmsd/ClarkeJEF96,
  author       = {Edmund M. Clarke and
                  Somesh Jha and
                  Reinhard Enders and
                  Thomas Filkorn},
  title        = {Exploiting Symmetry in Temporal Logic Model Checking},
  journal      = {Formal Methods Syst. Des.},
  volume       = {9},
  number       = {1/2},
  pages        = {77--104},
  year         = {1996},
  doi          = {10.1007/BF00625969},
}

@inproceedings{DBLP:conf/laser/ClarkeKNZ11,
  author       = {Edmund M. Clarke and
                  William Klieber and
                  Milos Nov{\'{a}}cek and
                  Paolo Zuliani},
  editor       = {Bertrand Meyer and
                  Martin Nordio},
  title        = {Model Checking and the State Explosion Problem},
  booktitle    = {Tools for Practical Software Verification, LASER, International Summer
                  School 2011, Elba Island, Italy, Revised Tutorial Lectures},
  series       = {Lecture Notes in Computer Science},
  volume       = {7682},
  pages        = {1--30},
  publisher    = {Springer},
  year         = {2011},
  doi          = {10.1007/978-3-642-35746-6_1},
}

@misc{Rocq,
  author    = {{The Rocq Development Team}},
  title     = {The {Rocq} Prover},
  publisher = {Zenodo},
  year      = {2026},
  url       = {https://doi.org/10.5281/zenodo.1003420}
}

@book{DBLP:books/sp/NipkowPW02,
  author       = {Tobias Nipkow and
                  Lawrence C. Paulson and
                  Markus Wenzel},
  title        = {{Isabelle/HOL} - {A} Proof Assistant for Higher-Order Logic},
  series       = {Lecture Notes in Computer Science},
  volume       = {2283},
  publisher    = {Springer},
  year         = {2002},
  doi          = {10.1007/3-540-45949-9},
  isbn         = {3-540-43376-7},
}

@inproceedings{DBLP:conf/cade/Moura021,
  author       = {Leonardo de Moura and
                  Sebastian Ullrich},
  editor       = {Andr{\'{e}} Platzer and
                  Geoff Sutcliffe},
  title        = {The {Lean} 4 Theorem Prover and Programming Language},
  booktitle    = {Automated Deduction - {CADE} 28 - 28th International Conference on
                  Automated Deduction, Virtual Event, July 12-15, 2021, Proceedings},
  series       = {Lecture Notes in Computer Science},
  volume       = {12699},
  pages        = {625--635},
  publisher    = {Springer},
  year         = {2021},
  doi          = {10.1007/978-3-030-79876-5_37},
}

@article{DBLP:journals/cacm/DeMilloLP79,
  author       = {Richard A. {DeMillo} and
                  Richard J. Lipton and
                  Alan J. Perlis},
  title        = {Social Processes and Proofs of Theorems and Programs},
  journal      = {Commun. {ACM}},
  volume       = {22},
  number       = {5},
  pages        = {271--280},
  year         = {1979},
  doi          = {10.1145/359104.359106},
}

@inproceedings{DBLP:conf/icse/MatichukMAJKS15,
  author       = {Daniel Matichuk and
                  Toby C. Murray and
                  June Andronick and
                  D. Ross Jeffery and
                  Gerwin Klein and
                  Mark Staples},
  editor       = {Antonia Bertolino and
                  Gerardo Canfora and
                  Sebastian G. Elbaum},
  title        = {Empirical Study Towards a Leading Indicator for Cost of Formal Software
                  Verification},
  booktitle    = {37th {IEEE/ACM} International Conference on Software Engineering,
                  {ICSE} 2015, Florence, Italy, May 16-24, 2015, Volume 1},
  pages        = {722--732},
  publisher    = {{IEEE} Computer Society},
  year         = {2015},
  doi          = {10.1109/ICSE.2015.85},
}

@article{DBLP:journals/pacmpl/KlimisCBNWD23,
  author       = {Vasileios Klimis and
                  Jack Clark and
                  Alan Baker and
                  David Neto and
                  John Wickerson and
                  Alastair F. Donaldson},
  title        = {Taking Back Control in an Intermediate Representation for {GPU} Computing},
  journal      = {Proc. {ACM} Program. Lang.},
  volume       = {7},
  number       = {{POPL}},
  pages        = {1740--1769},
  year         = {2023},
  doi          = {10.1145/3571253},
}

@article{DBLP:journals/tse/BettsCDDK18,
  author       = {Adam Betts and
                  Nathan Chong and
                  Pantazis Deligiannis and
                  Alastair F. Donaldson and
                  Jeroen Ketema},
  title        = {Implementing and Evaluating Candidate-Based Invariant Generation},
  journal      = {{IEEE} Trans. Software Eng.},
  volume       = {44},
  number       = {7},
  pages        = {631--650},
  year         = {2018},
  doi          = {10.1109/TSE.2017.2718516},
}

@techreport{MetamorphicTesting,
  title = {Metamorphic Testing: a New Approach for Generating Next Test Cases},
  author = {T.Y. Chen and S.C. Cheung and S.M. Yiu},
  institution = {Department of Computer Science, The Hong Kong University of Science and Technology},
  year = {1998},
  number = {HKUST-CS98-01}
}

@article{DBLP:journals/pacmpl/Lee-JonesWD26,
  author       = {James Lee{-}Jones and
                  John Wickerson and
                  Alastair F. Donaldson},
  title        = {Uniformity Analysis in the {WebGPU} Shading Language},
  journal      = {Proc. {ACM} Program. Lang.},
  volume       = {10},
  number       = {{PLDI}},
  pages        = {2109--2133},
  year         = {2026},
  doi          = {10.1145/3808331},
}

@misc{aria2026keystone,
  author       = {Alastair F. Donaldson and Marios Kogias and John Wickerson and Thomas Bourgeat and Nate Foster and Cl\'{e}ment Pit-Claudel},
  title        = {Keystone: verifying the {AI} inference stack},
  year         = {2026},
  howpublished = {\url{https://aria.org.uk/opportunity-spaces/trust-everything-everywhere/safeguarded-ai/funded-projects?tabId=applications&cardId=keystone}},
  note         = {Advanced Research and Invention Agency (ARIA),
                  accessed 13 October 2026}
}

@misc{yu2026randomisedformalisation,
  author       = {Alastair F. Donaldson},
  title        = {Randomised Generation Formalisation:
                  Proof that {GEN} = {RE}},
  year         = {2026},
  url          = {https://github.com/mc-imperial/RandomisedGenerationFormalisation},
}

@article{DBLP:journals/corr/abs-2607-11811,
  author       = {Pingshi Yu and
                  Chengsong Tan and
                  Nicolas Wu and
                  Alastair F. Donaldson},
  title        = {Complexity Theory of Randomised Testing},
  journal      = {CoRR},
  volume       = {abs/2607.11811},
  year         = {2026},
  doi          = {10.48550/ARXIV.2607.11811},
  eprinttype   = {arXiv},
  eprint       = {2607.11811},
}

@inproceedings{DBLP:conf/setss/Bowen14,
  author       = {Jonathan P. Bowen},
  editor       = {Zhiming Liu and
                  Zili Zhang},
  title        = {The {Z} Notation: Whence the Cause and Whither the Course?},
  booktitle    = {Engineering Trustworthy Software Systems - First International School,
                  {SETSS} 2014, Chongqing, China, September 8-13, 2014. Tutorial Lectures},
  series       = {Lecture Notes in Computer Science},
  volume       = {9506},
  pages        = {103--151},
  publisher    = {Springer},
  year         = {2014},
  doi          = {10.1007/978-3-319-29628-9_3},
}

@techreport{dijkstra1970notes,
  author      = {Edsger W. Dijkstra},
  title       = {Notes on Structured Programming},
  institution = {Technische Hogeschool Eindhoven},
  number      = {70-WSK-03},
  year        = {1970},
  month       = apr,
  edition     = {Second},
  note        = {EWD249},
  url         = {https://www.cs.utexas.edu/~EWD/ewd02xx/EWD249.PDF}
}

@book{holzmann2004spin,
  author    = {Gerard J. Holzmann},
  title     = {The {SPIN} Model Checker: Primer and Reference Manual},
  publisher = {Addison-Wesley},
  year      = {2004},
  isbn      = {9780321228628}
}

@inproceedings{DBLP:conf/ifm/BallCLR04,
  author       = {Thomas Ball and
                  Byron Cook and
                  Vladimir Levin and
                  Sriram K. Rajamani},
  editor       = {Eerke A. Boiten and
                  John Derrick and
                  Graeme Smith},
  title        = {{SLAM} and {Static Driver Verifier}: Technology Transfer of Formal Methods
                  inside {Microsoft}},
  booktitle    = {Integrated Formal Methods, 4th International Conference, {IFM} 2004,
                  Canterbury, UK, April 4-7, 2004, Proceedings},
  series       = {Lecture Notes in Computer Science},
  volume       = {2999},
  pages        = {1--20},
  publisher    = {Springer},
  year         = {2004},
  doi          = {10.1007/978-3-540-24756-2_1},
}

@inproceedings{DBLP:conf/sosp/KleinEHACDEEKNSTW09,
  author       = {Gerwin Klein and
                  Kevin Elphinstone and
                  Gernot Heiser and
                  June Andronick and
                  David A. Cock and
                  Philip Derrin and
                  Dhammika Elkaduwe and
                  Kai Engelhardt and
                  Rafal Kolanski and
                  Michael Norrish and
                  Thomas Sewell and
                  Harvey Tuch and
                  Simon Winwood},
  editor       = {Jeanna Neefe Matthews and
                  Thomas E. Anderson},
  title        = {se{L}4: formal verification of an {OS} kernel},
  booktitle    = {Proceedings of the 22nd {ACM} Symposium on Operating Systems Principles
                  2009, {SOSP} 2009, Big Sky, Montana, USA, October 11-14, 2009},
  pages        = {207--220},
  publisher    = {{ACM}},
  year         = {2009},
  doi          = {10.1145/1629575.1629596},
}

@article{DBLP:journals/cacm/Leroy09,
  author       = {Xavier Leroy},
  title        = {Formal verification of a realistic compiler},
  journal      = {Commun. {ACM}},
  volume       = {52},
  number       = {7},
  pages        = {107--115},
  year         = {2009},
  doi          = {10.1145/1538788.1538814},
}

@inproceedings{DBLP:conf/tacas/JhalaM06,
  author       = {Ranjit Jhala and
                  Kenneth L. McMillan},
  editor       = {Holger Hermanns and
                  Jens Palsberg},
  title        = {A Practical and Complete Approach to Predicate Refinement},
  booktitle    = {Tools and Algorithms for the Construction and Analysis of Systems,
                  12th International Conference, {TACAS} 2006 Held as Part of the Joint
                  European Conferences on Theory and Practice of Software, {ETAPS} 2006,
                  Vienna, Austria, March 25 - April 2, 2006, Proceedings},
  series       = {Lecture Notes in Computer Science},
  volume       = {3920},
  pages        = {459--473},
  publisher    = {Springer},
  year         = {2006},
  doi          = {10.1007/11691372_33},
}

@article{DBLP:journals/toplas/BettsCDKQTW15,
  author       = {Adam Betts and
                  Nathan Chong and
                  Alastair F. Donaldson and
                  Jeroen Ketema and
                  Shaz Qadeer and
                  Paul Thomson and
                  John Wickerson},
  title        = {The Design and Implementation of a Verification Technique for {GPU}
                  Kernels},
  journal      = {{ACM} Trans. Program. Lang. Syst.},
  volume       = {37},
  number       = {3},
  pages        = {10:1--10:49},
  year         = {2015},
  doi          = {10.1145/2743017},
}

@inproceedings{DBLP:conf/icst/DonaldsonSTU24,
  author       = {Alastair F. Donaldson and
                  Dilan Sheth and
                  Jean{-}Baptiste Tristan and
                  Alex Usher},
  title        = {Randomised Testing of the Compiler for a Verification-Aware Programming
                  Language},
  booktitle    = {{IEEE} Conference on Software Testing, Verification and Validation,
                  {ICST} 2024, Toronto, ON, Canada, May 27-31, 2024},
  pages        = {407--418},
  publisher    = {{IEEE}},
  year         = {2024},
  doi          = {10.1109/ICST60714.2024.00044},
}

@inproceedings{DBLP:conf/issta/IrfanPRRT22,
  author       = {Ahmed Irfan and
                  Sorawee Porncharoenwase and
                  Zvonimir Rakamaric and
                  Neha Rungta and
                  Emina Torlak},
  editor       = {Sukyoung Ryu and
                  Yannis Smaragdakis},
  title        = {Testing {Dafny} (experience paper)},
  booktitle    = {{ISSTA} '22: 31st {ACM} {SIGSOFT} International Symposium on Software
                  Testing and Analysis, Virtual Event, South Korea, July 18 - 22, 2022},
  pages        = {556--567},
  publisher    = {{ACM}},
  year         = {2022},
  doi          = {10.1145/3533767.3534382},
}

@article{DBLP:journals/pacmpl/0001024a,
  author       = {Chengyu Zhang and
                  Zhendong Su},
  title        = {{SMT2Test}: From {SMT} Formulas to Effective Test Cases},
  journal      = {Proc. {ACM} Program. Lang.},
  volume       = {8},
  number       = {{OOPSLA2}},
  pages        = {222--245},
  year         = {2024},
  doi          = {10.1145/3689719},
}

@article{DBLP:journals/pacmpl/DonaldsonELT17,
  author       = {Alastair F. Donaldson and
                  Hugues Evrard and
                  Andrei Lascu and
                  Paul Thomson},
  title        = {Automated testing of graphics shader compilers},
  journal      = {Proc. {ACM} Program. Lang.},
  volume       = {1},
  number       = {{OOPSLA}},
  pages        = {93:1--93:29},
  year         = {2017},
  doi          = {10.1145/3133917},
}

@article{DBLP:journals/pacmpl/RiggerS20,
  author       = {Manuel Rigger and
                  Zhendong Su},
  title        = {Finding bugs in database systems via query partitioning},
  journal      = {Proc. {ACM} Program. Lang.},
  volume       = {4},
  number       = {{OOPSLA}},
  pages        = {211:1--211:30},
  year         = {2020},
  doi          = {10.1145/3428279},
}

@inproceedings{DBLP:conf/pldi/WintererZS20,
  author       = {Dominik Winterer and
                  Chengyu Zhang and
                  Zhendong Su},
  editor       = {Alastair F. Donaldson and
                  Emina Torlak},
  title        = {Validating {SMT} solvers via semantic fusion},
  booktitle    = {Proceedings of the 41st {ACM} {SIGPLAN} International Conference on
                  Programming Language Design and Implementation, {PLDI} 2020, London,
                  UK, June 15-20, 2020},
  pages        = {718--730},
  publisher    = {{ACM}},
  year         = {2020},
  doi          = {10.1145/3385412.3385985},
}

@inproceedings{DBLP:conf/pldi/LeAS14,
  author       = {Vu Le and
                  Mehrdad Afshari and
                  Zhendong Su},
  editor       = {Michael F. P. O'Boyle and
                  Keshav Pingali},
  title        = {Compiler validation via equivalence modulo inputs},
  booktitle    = {{ACM} {SIGPLAN} Conference on Programming Language Design and Implementation,
                  {PLDI} '14, Edinburgh, United Kingdom - June 09 - 11, 2014},
  pages        = {216--226},
  publisher    = {{ACM}},
  year         = {2014},
  doi          = {10.1145/2594291.2594334},
}

@inproceedings{DBLP:conf/pldi/LidburyLCD15,
  author       = {Christopher Lidbury and
                  Andrei Lascu and
                  Nathan Chong and
                  Alastair F. Donaldson},
  editor       = {David Grove and
                  Stephen M. Blackburn},
  title        = {Many-core compiler fuzzing},
  booktitle    = {Proceedings of the 36th {ACM} {SIGPLAN} Conference on Programming
                  Language Design and Implementation, Portland, OR, USA, June 15-17,
                  2015},
  pages        = {65--76},
  publisher    = {{ACM}},
  year         = {2015},
  doi          = {10.1145/2737924.2737986},
}

@inproceedings{DBLP:conf/ecoop/DonaldsonET19,
  author       = {Alastair F. Donaldson and
                  Hugues Evrard and
                  Paul Thomson},
  editor       = {Robert Hirschfeld and
                  Tobias Pape},
  title        = {Putting Randomized Compiler Testing into Production (Experience Report)},
  booktitle    = {34th European Conference on Object-Oriented Programming, {ECOOP} 2020,
                  Berlin, Germany (Virtual Conference), November 15-17, 2020},
  series       = {LIPIcs},
  volume       = {166},
  pages        = {22:1--22:29},
  publisher    = {Schloss Dagstuhl - Leibniz-Zentrum f{\"{u}}r Informatik},
  year         = {2020},
  doi          = {10.4230/LIPICS.ECOOP.2020.22},
}

@inproceedings{DBLP:conf/pldi/YangCER11,
  author       = {Xuejun Yang and
                  Yang Chen and
                  Eric Eide and
                  John Regehr},
  editor       = {Mary W. Hall and
                  David A. Padua},
  title        = {Finding and understanding bugs in {C} compilers},
  booktitle    = {Proceedings of the 32nd {ACM} {SIGPLAN} Conference on Programming
                  Language Design and Implementation, {PLDI} 2011, San Jose, CA, USA,
                  June 4-8, 2011},
  pages        = {283--294},
  publisher    = {{ACM}},
  year         = {2011},
  doi          = {10.1145/1993498.1993532},
}

@inproceedings{DBLP:conf/nfm/CuoqMPPRYY12,
  author       = {Pascal Cuoq and
                  Benjamin Monate and
                  Anne Pacalet and
                  Virgile Prevosto and
                  John Regehr and
                  Boris Yakobowski and
                  Xuejun Yang},
  editor       = {Alwyn Goodloe and
                  Suzette Person},
  title        = {Testing Static Analyzers with Randomly Generated Programs},
  booktitle    = {{NASA} Formal Methods - 4th International Symposium, {NFM} 2012, Norfolk,
                  VA, USA, April 3-5, 2012. Proceedings},
  series       = {Lecture Notes in Computer Science},
  volume       = {7226},
  pages        = {120--125},
  publisher    = {Springer},
  year         = {2012},
  doi          = {10.1007/978-3-642-28891-3_12},
}

\end{document}